\documentclass[aps,prd,superscriptaddress,notitlepage,showpacs,showkeys,10pt]{revtex4-1}

\usepackage{graphicx}
\usepackage{amsmath}
\usepackage{amssymb}
\usepackage{color}
\usepackage{bm}

\definecolor{purple}{rgb}{0.8,0,0.6}
\definecolor{darkgreen}{rgb}{0.00,0.6,0.00}

\begin{document}
\title{Collective excitations in Weyl semimetals in the hydrodynamic regime}
\date{June 11, 2018}

\author{P.~O.~Sukhachov}
\affiliation{Department of Applied Mathematics, Western University, London, Ontario, Canada N6A 5B7}

\author{E.~V.~Gorbar}
\affiliation{Department of Physics, Taras Shevchenko National Kiev University, Kiev, 03680, Ukraine}
\affiliation{Bogolyubov Institute for Theoretical Physics, Kiev, 03680, Ukraine}

\author{I.~A.~Shovkovy}
\affiliation{College of Integrative Sciences and Arts, Arizona State University, Mesa, Arizona 85212, USA}
\affiliation{Department of Physics, Arizona State University, Tempe, Arizona 85287, USA}

\author{V.~A.~Miransky}
\affiliation{Department of Applied Mathematics, Western University, London, Ontario, Canada N6A 5B7}

\begin{abstract}
The spectrum of collective excitations in Weyl materials is studied by using consistent
hydrodynamics. The corresponding framework includes the vortical and chiral anomaly
effects, as well as the dependence on the separations between the Weyl nodes in energy $b_0$
and momentum $\mathbf{b}$. The latter are introduced via the Chern--Simons contributions to the electric
current and charge densities in Maxwell's equations. It is found that, even in the absence of a
background magnetic field, certain collective excitations (e.g. the helicon-like modes and the anomalous Hall waves) are
strongly affected by the chiral shift $\mathbf{b}$. In a background magnetic field,
the existence of the distinctive longitudinal and transverse anomalous Hall waves with a linear dispersion relation is predicted.
They originate from the oscillations of the electric charge density and electromagnetic fields,
in which different components of the fields are connected via the anomalous
Hall effect in Weyl semimetals.
\end{abstract}

\maketitle

\section{Introduction}
\label{sec:introduction}

Collective excitations play an important role in all kinds of plasmas \cite{Krall,Landau:t10,Maxfield,Kaner},
ranging from the nonrelativistic electron-hole ones in solids \cite{Maxfield,Kaner} to various relativistic plasmas created in the
heavy-ion collisions \cite{Kharzeev:2008-Nucl,Kharzeev:2016}, the early Universe \cite{Kronberg, Durrer},
and astrophysics \cite{Kouveliotou:1999}. With the discovery of graphene, whose quasiparticles
are described by the two-dimensional (2D) relativistic-like Dirac equation (for reviews, see, e.g. Refs.~\cite{Neto-Geim:rev,Sarma:rev,Katsnelson:rev}),
it became clear that relativistic-like plasmas could be also realized and studied in condensed matter
systems. Furthermore, the discovery of Dirac and Weyl semimetals with three-dimensional (3D) chiral quasiparticles (for
recent reviews, see Refs.~\cite{Yan-Felser:2017-Rev,Hasan-Huang:2017-Rev,Armitage-Vishwanath:2017-Rev})
made possible the study of relativistic phenomena in well-controlled table-top experiments.

The defining feature of Weyl semimetals is that the opposite chirality Weyl nodes are separated in energy and/or momentum.
This is in contrast to Dirac semimetals, where the location of such nodes
coincides.
The separation between the Weyl nodes is quantified by a pseudo-scalar
parameter $b_0$ and an axial vector $\mathbf{b}$ (the latter is also known as the chiral shift) that break the
parity-inversion (PI) and time-reversal (TR) symmetries, respectively. In addition, an imbalance between the number densities of opposite
chirality carriers could be induced. These chiral asymmetries, in turn, lead to qualitatively new effects including unusual properties of collective excitations in the plasma.

The spectrum of collective excitations in chiral relativistic plasmas can be studied
by using any approach that takes into account the chiral nature of quasiparticles and accounts for
the chiral anomaly~\cite{Adler,Bell-Jackiw}. Among the most popular of them are the chiral kinetic theory (CKT)
\cite{Son:2012wh,Stephanov:2012ki,Son,Chen:2014cla,Manuel:2014dza} and the chiral hydrodynamics
\cite{Son:2009tf,Sadofyev:2010pr,Neiman:2010zi,Hidaka:2017auj}. The kinetic theory allows one to investigate a
wide range of collective excitations with wavelengths smaller or equal to the particle mean free path
$\ell_{\rm mfp}$, provided the latter is much larger than the average interparticle distance. The
use of the chiral hydrodynamics, on the other hand, is limited to the length scales at which the
assumption of local equilibrium is satisfied, i.e. to collective excitations with wavelengths much
longer than $\ell_{\rm mfp}$ (or, equivalently, sufficiently small wave vectors, i.e. $k \ll 1/\ell_{\rm mfp}$).

The potential relevance of hydrodynamics for the description of the electron transport in solids
was understood long time ago by Gurzhi \cite{Gurzhi,Gurzhi-effect}.
The corresponding
framework is applicable only when the electron fluid is well defined and decoupled from the ion
lattice, i.e. when the rate of the electron-electron scattering is much higher than the rates of
electron scatterings on phonons and impurities. These conditions are usually met only in a finite
window of temperatures, limited by the growing scattering rates on impurities at small
and on phonons at high temperature. In the case of graphene, for example, such a hydrodynamic
window was discussed theoretically in Refs.~\cite{Levitov:2017,Ho-Adam:2017} and observed
experimentally in Refs.~\cite{Crossno,Ghahari}. It is interesting to mention that the viscous
hydrodynamic flow can provide a higher conduction than the ballistic transport \cite{Levitov:2017}.
Recently, the hydrodynamic regime was also experimentally probed in Weyl semimetals \cite{Gooth:2017}.
The hydrodynamical nature of the transport observations is supported by a characteristic dependence of the electrical resistivity on
the constriction channel width, as well as by a strong violation of the Wiedemann--Franz law with
the lowest value of the Lorenz number ever reported \cite{Gooth:2017}.

Theoretically, the hydrodynamic approach was already used to describe the negative magnetoresistance
\cite{Landsteiner:2014vua,Lucas:2016omy} and thermoelectric transport \cite{Lucas:2016omy}
in Weyl semimetals. However, the corresponding framework lacked any information
on the separation between the Weyl nodes. It is worth reminding that such a critical information is also missing in the conventional
chiral kinetic theory \cite{Son:2012wh,Stephanov:2012ki,Son,Chen:2014cla,Manuel:2014dza}
and is known to cause a serious problem. Indeed,
this theory naively predicts an anomalous local nonconservation of the electric charge when both electromagnetic
and strain-induced pseudoelectromagnetic fields are applied to the system \cite{Pikulin}. As we argued
in Ref.~\cite{Gorbar:2016ygi}, the problem is resolved in the framework of the consistent chiral kinetic
theory, in which the electric charge and current densities include the topological Chern--Simons contributions
\cite{Landsteiner:2016} (also known as the Bardeen--Zumino terms \cite{Bardeen:1969md,Bardeen:1984pm}
in high energy physics). These contributions introduce the needed dependence on the energy separation $b_0$
and the chiral shift $\mathbf{b}$ that are critical for
reproducing the correct chiral magnetic effect (CME) in equilibrium \cite{Franz,Basar:2014,Landsteiner:2016} and the anomalous Hall
effect (AHE) \cite{Ran,Burkov:2011ene,Grushin-AHE,Goswami,Burkov-AHE:2014} in Weyl semimetals. By noting that the hydrodynamic equations
can be obtained by averaging the corresponding kinetic ones (see, e.g. Refs.~\cite{Landau:t10,Huang-book}),
we recently argued \cite{Gorbar:2017vph} that, in the consistent hydrodynamics (CHD),
similar Chern--Simons contributions should be added to the electric current and charge densities in
Maxwell's equations. On the other hand, these terms are absent in the Euler equation and the energy conservation relation for the electron fluid.

Within the CKT approach, it was already demonstrated that the plasmons \cite{Gorbar:2016sey} and
helicons \cite{Pellegrino,Gorbar:2016vvg} are affected by the energy and momentum separations
between the Weyl nodes. At nonzero electric chemical
potential, for instance, the effect of the chiral shift is similar to that of a background magnetic field:
it lifts the degeneracy of plasmon frequencies and mixes the longitudinal modes with the transverse
ones. Also, as shown in Ref.~\cite{Pellegrino}, the component of the chiral shift along the direction
of an external magnetic field modifies the effective helicon mass. It is reasonable to expect, therefore, that
the collective excitations in the hydrodynamic regime are also affected by $b_0$ and $\mathbf{b}$.
The main goal of this paper is to investigate the corresponding spectrum of collective modes
in detail by using the CHD. (The case of transverse collective excitations was
briefly reported in Ref.~\cite{Gorbar:2017vph}.)

The paper is organized as follows. In Sec.~\ref{sec:model}, we review the key features of the CHD
in Weyl materials \cite{Gorbar:2017vph}. The linearized hydrodynamic equations together
with Maxwell's equations are discussed in Sec.~\ref{sec:Minimal-WH-2}. The
collective excitations in the absence of a background magnetic field are studied in Sec.~\ref{sec:WMHD-B0}.
The longitudinal and transverse collective modes, defined with respect to the direction of the magnetic field, are
analyzed in Secs.~\ref{sec:WMHD-B-kz} and \ref{sec:WMHD-B-kx}, respectively. In Sec.~\ref{sec:Discussions},
we discuss the range of validity of the obtained results and the experimental techniques that could be
used to test the predictions. The summary of our study is given in Sec.~\ref{sec:Summary}.
Technical details, including the explicit expressions for the transport coefficients and the frequencies of some collective modes are given in Appendix~\ref{sec:app-formulas}. Throughout this paper, we set the Boltzmann constant $k_B=1$.

\section{Consistent hydrodynamic theory}
\label{sec:model}

In this section, we review the CHD
in Weyl semimetals derived in Ref.~\cite{Gorbar:2017vph}.
The corresponding equations take into account the explicit breaking of the Galilean invariance by the ion lattice
and include the effects of the vorticity and the chiral anomaly. (Note that the CHD also describes Dirac
semimetals if one sets $b_0=0$ and $\mathbf{b}=\mathbf{0}$.)

Let us begin with the electric and chiral charge continuity relations
\begin{eqnarray}
\label{model-J-conserv-eq}
\partial_t \rho +\left(\bm{\nabla}\cdot\mathbf{J}\right) &=& 0, \\
\label{model-J5-conserv-eq}
\partial_t \rho_5 +\left(\bm{\nabla}\cdot\mathbf{J}_5\right) &=& -\frac{e^3 (\mathbf{E}\cdot\mathbf{B})}{2\pi^2 \hbar^2 c},
\end{eqnarray}
where $\rho$ and $\rho_{5}$ denote the \emph{total} electric and chiral charge densities, respectively.
The \emph{total} electric and chiral current densities are denoted as $\mathbf{J}$ and $\mathbf{J}_5$,
respectively. Their explicit expressions read \cite{Gorbar:2017vph}
\begin{eqnarray}
\label{model-rho-def}
\rho &=& -e n +\frac{\sigma^{(B)} \left(\mathbf{B}\cdot\mathbf{u}\right)}{3v_F^2} + \frac{5c \sigma^{(\epsilon, u)} \left(\mathbf{B}\cdot\bm{\omega}\right)}{v_F^2}
+\rho_{\text{{\tiny CS}}},\\
\label{model-rho5-def}
\rho_5 &=& -en_{5} +\frac{\sigma^{(B)}_5 \left(\mathbf{B}\cdot\mathbf{u}\right) }{3v_F^2},
\end{eqnarray}
and
\begin{eqnarray}
\label{model-J-def}
\mathbf{J} &=& -en\mathbf{u} +\sigma^{(V)}\bm{\omega} +\sigma^{(B)}\mathbf{B}+\frac{c\sigma^{(B)}\left[\mathbf{E}\times\mathbf{u}\right]}{3 v_F^2}  + \frac{5c^2\sigma^{(\epsilon, u)}\left[\mathbf{E}\times\bm{\omega}\right]}{v_F^2}
-\frac{\left[\mathbf{u}\times\bm{\nabla}\right]\sigma^{(V)}}{3} +\frac{\left[\bm{\nabla}\times\bm{\omega}\right] \sigma^{(\epsilon, V)}}{2} +\mathbf{J}_{\text{{\tiny CS}}},\\
\label{model-J5-def}
\mathbf{J}_5 &=& -en_{5}\mathbf{u} +\sigma^{(V)}_5\bm{\omega} +\sigma^{(B)}_5\mathbf{B}+\frac{c\sigma^{(B)}_5\left[\mathbf{E}\times\mathbf{u}\right]}{3v_F^2} -\frac{\left[\mathbf{u}\times\bm{\nabla}\right]\sigma^{(V)}_5}{3}  +\frac{\left[\bm{\nabla}\times\bm{\omega}\right] \sigma^{(\epsilon, V)}_5}{2}.
\end{eqnarray}
Here $e$ is the absolute value of the electron charge, $\mathbf{u}$ is the fluid velocity,
$\bm{\omega}=\left[\bm{\nabla}\times\mathbf{u}\right]/2$ is the vorticity, $v_F$ is the Fermi velocity,
and $c$ is the speed of light.
Here $n$ and $n_5$ are the fermion and the chiral fermion number densities, respectively. The explicit expressions for the anomalous transport coefficients are given in Appendix~\ref{sec:app-formulas-coeff}.

In addition to the matter parts of the charge and current densities, there are also the
topological Chern--Simons contributions $\rho_{\text{{\tiny CS}}}$ and $\mathbf{J}_{\text{{\tiny CS}}}$
\cite{Bardeen:1969md,Bardeen:1984pm,Landsteiner:2013sja,Landsteiner:2016,Gorbar:2016ygi,Gorbar:2017-Bardeen,Gorbar:2017vph}
given by
\begin{eqnarray}
\label{model-CS-charge}
\rho_{\text{{\tiny CS}}} &=&- \frac{e^3 (\mathbf{b}\cdot\mathbf{B})}{2\pi^2\hbar^2c^2},\\
\label{model-CS-current}
\mathbf{J}_{\text{{\tiny CS}}} &=&-\frac{e^3b_0 \mathbf{B}}{2\pi^2\hbar^2c} + \frac{e^3\left[\mathbf{b}\times\mathbf{E}\right]}{2\pi^2\hbar^2c}.
\end{eqnarray}
Here it is important to emphasize that the above Chern--Simons contributions affect Maxwell's equations in Weyl semimetals and, in fact, are the only source of the dependence on the energy separation
$b_0$ and the chiral shift $\mathbf{b}$ in the CHD.

A few comments regarding the charge and current densities are in order here. First, we note that the charge densities
are modified by the fluid velocity and vorticity. Next, as expected, the conventional CME and the chiral vortical effect (CVE)
\cite{Chen:2014cla}, as well as their chiral counterparts are reproduced in the current densities. Among the vorticity-related contributions,
there are terms proportional to the cross product of the fluid velocity and gradients of the thermodynamic variables,
as well as curls of vorticity. The terms $\propto (\mathbf{B}\cdot\mathbf{u})$ could be considered as anomalous charge
inflows. There is also a term $\propto (\mathbf{B}\cdot\bm{\omega})$ that is structurally similar, but its coefficient
does not depend on the chemical potentials. In the current densities, we also included the terms induced by the
simultaneous presence of the fluid velocity and the electric field. While they resemble the contributions obtained via the
usual Lorentz transformation of the terms with the magnetic field, their coefficients are different. This should not be
surprising, however, since there is no true Lorentz invariance in Weyl and Dirac semimetals.

As is well known, the hydrodynamic equations can be obtained from the corresponding kinetic ones
\cite{Landau:t10,Huang-book} by assuming that the deviations from the state of local equilibrium are small.
By making use of the consistent CKT \cite{Gorbar:2016ygi}
in the relaxation-time approximation, we derived the
Euler equation and the energy conservation relation for the electron fluid in Weyl and Dirac semimetals in
Ref.~\cite{Gorbar:2017vph} (for the details of the derivation, see the Supplemental Material in Ref.~\cite{Gorbar:2017vph}).

The explicit form of the Euler equation reads
\begin{eqnarray}
\label{model-Euler}
&&\frac{1}{v_F}\partial_t \left[\frac{ w  \mathbf{u}}{v_F}+\sigma^{(\epsilon,B)}\mathbf{B}
+\frac{\hbar \bm{\omega} n_{5}}{2}\right] +\bm{\nabla}_{\mathbf{r}}P + \frac{4}{15v_F} \left[\sum_{j=1}^3 \mathbf{B}_j \bm{\nabla}_{\mathbf{r}}\mathbf{u}_j + (\mathbf{B}\cdot\bm{\nabla}_{\mathbf{r}})\mathbf{u} +\mathbf{B} (\bm{\nabla}_{\mathbf{r}}\cdot \mathbf{u})\right] \sigma^{(\epsilon, B)} \nonumber\\
&&+ \frac{c \left[\bm{\nabla}_{\mathbf{r}}\times\mathbf{E}\right] \sigma^{(\epsilon, B)}}{3v_F}
+\frac{2}{3v_F}\sum_{j=1}^3 \mathbf{u}_j\bm{\nabla}_{\mathbf{r}}\mathbf{B}_j \sigma^{(\epsilon, B)}
-\frac{4\sigma^{(\epsilon, B)}}{15v_F} \left[\sum_{j=1}^3 \mathbf{u}_j\bm{\nabla}_{\mathbf{r}}\mathbf{B}_j +(\mathbf{u}\cdot\bm{\nabla}_{\mathbf{r}})\mathbf{B}\right]
+\frac{5\sigma^{(\epsilon, u)}\bm{\nabla}_{\mathbf{r}}B^2}{2} \nonumber\\
&&+ \left[(\mathbf{B}\cdot\bm{\nabla}_{\mathbf{r}})\bm{\omega} + \sum_{j=1}^3 \mathbf{B}_j\bm{\nabla}_{\mathbf{r}} \bm{\omega}_j  \right] \frac{\sigma^{(\epsilon, V)}}{5c} -\sum_{j=1}^3 \frac{\sigma^{(\epsilon, V)}\bm{\omega}_j \bm{\nabla}_{\mathbf{r}}\mathbf{B}_j}{2c}
-\frac{\sigma^{(\epsilon, V)}}{5c} \left[\sum_{j=1}^3\bm{\omega}_j\bm{\nabla}_{\mathbf{r}}\mathbf{B}_j + (\bm{\omega}\cdot\bm{\nabla}_{\mathbf{r}})\mathbf{B} \right] \nonumber\\
&& =-en\mathbf{E} +\frac{1}{c}\left[\mathbf{B}\times \left(en\mathbf{u}
-\frac{\sigma^{(V)}\bm{\omega}}{3}
\right)\right] +\frac{\sigma^{(B)} \mathbf{u} (\mathbf{E}\cdot\mathbf{B})}{3v_F^2} +\frac{5c \sigma^{(\epsilon, u)} (\mathbf{E}\cdot\mathbf{B})\bm{\omega}}{v_F}
 -\frac{ w  \mathbf{u}}{v_F^2\tau} - \frac{\hbar \bm{\omega} n_5}{2 v_F \tau},
\end{eqnarray}
where $B\equiv|\mathbf{B}|$, $w=\epsilon+P$ is the enthalpy density, $\epsilon$ is the energy density, and $P$ is the pressure.
In this equation, we used the convention that the derivatives apply to all quantities standing to their right.
Further, $\tau$ denotes the relaxation time connected with the intravalley (chirality preserving) scattering,
which is the dominant dissipation mechanism. In this connection, we note that the relaxation time $\tau_5$
for the chirality-flipping (intervalley) processes is usually much larger \cite{Zhang-Xiu:2015}
and, therefore, can be ignored.

A rather complicated form of the Euler equation (\ref{model-Euler}) for the electron fluid in Weyl and Dirac semimetals
is the result of including many anomalous terms and the vorticity effects. Before proceeding further,
therefore, it is instructive to discuss some of its most interesting features. Firstly, as we noted in Ref.~\cite{Gorbar:2017vph},
the Chern--Simons current density $\mathbf{J}_{\text{\tiny CS}}$ does not contribute directly to the Euler
equation. Secondly, the dissipation effects are captured primarily by the term $\propto \mathbf{u}/\tau$,
which is a distinctive feature of an electron fluid in solids \cite{Gurzhi,Gurzhi-effect}. It originates from the electron scattering
on phonons and/or impurities and breaks explicitly the Galilean symmetry. The breaking of the Galilean symmetry
should not be surprising because there is a preferred coordinate system in which the lattice ions are stationary.
It is worth noting that the term $\propto \mathbf{u}/\tau$ is responsible for the usual Ohm's
law in the steady state of the electron liquid. Indeed, at $\mathbf{B}= \mathbf{0}$, the steady state is reached
when the right-hand side of the Euler equation (\ref{model-Euler}) vanishes, i.e. when
$\mathbf{u}_{\rm ave} = - e n \tau v_F^2\mathbf{E}/w$. After taking into account the
hydrodynamic definition of the electric current density $\mathbf{J} = -en \mathbf{u}_{\rm ave}$, the latter
reproduces the conventional Ohm's law. Thirdly, we note that the Lorentz force term on the right-hand
side of the Euler equation contains an extra factor of $1/3$ in the CVE current. Lastly, all
terms $\propto\left(\mathbf{E}\cdot\mathbf{B}\right)$ on the right-hand side of Eq.~(\ref{model-Euler})
are related to the chiral anomaly.

In addition to the Euler equation, we also need the following energy conservation relation \cite{Gorbar:2017vph}:
\begin{eqnarray}
\label{model-energy}
&&\partial_t \epsilon
+(\bm{\nabla}_{\mathbf{r}}\cdot\mathbf{u})  w  +\sigma^{(\epsilon, u)} \left[\sum_{i=1}^3\mathbf{B}_i \left(\mathbf{B}\cdot\bm{\nabla}_{\mathbf{r}}\right)\mathbf{u}_i -2B^2(\bm{\nabla}_{\mathbf{r}}\cdot\mathbf{u})\right] -\frac{2c\left(\mathbf{E}\cdot\left[\bm{\nabla}_{\mathbf{r}}\times\mathbf{u}\right]\right)\sigma^{(\epsilon,B)}}{3v_F}\nonumber\\
&&-5c\sigma^{(\epsilon, u)}\left(\mathbf{E}\cdot\left[\bm{\nabla}_{\mathbf{r}}\times\mathbf{B}\right]\right)
+v_F\left(\mathbf{B}\cdot\bm{\nabla}_{\mathbf{r}}\right)\sigma^{(\epsilon, B)}
-2\sigma^{(\epsilon, u)}
\left[\left(\mathbf{u}\cdot\bm{\nabla}_{\mathbf{r}}\right)B^2 -3\sum_{i=1}^3\mathbf{u}_i\left(\mathbf{B}\cdot\bm{\nabla}_{\mathbf{r}}\right)\mathbf{B}_i\right]
\nonumber\\
&& +\frac{\hbar v_F (\bm{\nabla}_{\mathbf{r}}\cdot\bm{\omega})n_5}{2} - \frac{\left(\mathbf{E}\cdot\left[\bm{\nabla}_{\mathbf{r}} \times \bm{\omega}\right]\right)\sigma^{(\epsilon, V)}}{2} =- \mathbf{E}\cdot \left(en\mathbf{u}-\sigma^{(B)}\mathbf{B} -\frac{\sigma^{(V)}\bm{\omega}}{3}\right).
\end{eqnarray}
As in the Euler equation, the electrical force term on the right-hand side of Eq.~(\ref{model-energy}) includes
an extra factor of $1/3$ in the CVE current. Also, there are no topological Chern--Simons contributions.

In order to obtain a self-consistent framework for the description of a chiral electron fluid in Weyl
semimetals, Eqs.~(\ref{model-J-conserv-eq}), (\ref{model-J5-conserv-eq}), (\ref{model-Euler}), and
(\ref{model-energy}) should be supplemented by the standard Maxwell's equations. It is worthwhile to
note that Gauss's law includes the \emph{total} electric charge density $\rho$ due to electrons given
by Eq.~(\ref{model-rho-def}) and the background charge density $\rho_{\rm b}$ (its value will be specified in the next section).
Similarly, Ampere's law contains the \emph{total} electric current density given by Eq.~(\ref{model-J-def}).

\section{Linearized hydrodynamics equations}
\label{sec:Minimal-WH-2}

In the study of collective modes, it is justified to assume that the deviations of the local thermodynamic
parameters from their equilibrium values always remain small. Therefore, a simpler linearized form of
the hydrodynamic equations is sufficient. In this section, we present the corresponding linearized equations
of the CHD.

The state of \emph{global} equilibrium in the chiral electron fluid is defined by the values of
the electric chemical potential $\mu_{0}$, the chiral chemical potential $\mu_{5,0}$, and temperature $T_0$.
The corresponding energy and electric charge densities are given by the following standard expressions:
\begin{eqnarray}
\label{Minimal-WH-2-equilibrium-be}
\epsilon_{0} &=& \frac{\mu^4_{0}+6\mu^2_{0}\mu_{5,0}^2+\mu_{5,0}^4}{4\pi^2\hbar^3v_F^3} +\frac{T^2_0(\mu^2_{0}+\mu_{5,0}^2)}{2\hbar^3v_F^3} +\frac{7\pi^2T^4_0}{60\hbar^3v_F^3},\\
\rho_0 &=&
-e n_0 = -e\frac{\mu_{0}\left(\mu^2_{0}+3\mu^2_{5,0}+\pi^2T^2_0\right)}{3\pi^2 \hbar^3 v_F^3 }.
\label{Minimal-WH-2-equilibrium-ee}
\end{eqnarray}
In addition, the pressure and the enthalpy density are $P_0=\epsilon_0/3$ and $w_0=4\epsilon_0/3$, respectively.
The expression for the chiral charge density $-e n_{5,0}$ is similar to that in Eq.~(\ref{Minimal-WH-2-equilibrium-ee}), but with the electric and chiral chemical potentials interchanged, $\mu_0\leftrightarrow\mu_{5,0}$.

In an electrically neutral sample, the \emph{total} equilibrium electron charge density $\rho_0$ defined with respect to the Weyl nodes
should be compensated by the background charge density due to ions in the material, i.e.
$\rho_0 +\rho_{\rm b} =0$. In the absence of a background magnetic field, this condition implies
\begin{eqnarray}
\label{Minimal-WH-2-compensation-B0}
e\frac{\mu_{0}\left(\mu^2_{0}+3\mu^2_{5,0}+\pi^2T^2_{0}\right)}{3\pi^2 \hbar^3 v_F^3} -\rho_{\rm b}=0,
\end{eqnarray}
where $\mu_{0}$ is the equilibrium value of the electric chemical potential at $B_0=0$.

In the presence of the background magnetic field $\mathbf{B}_0\parallel\hat{\mathbf{z}}$, the electric current should be absent in
the global equilibrium state, i.e.
\begin{equation}
\label{Minimal-WH-2-J0-compensation}
\mathbf{J}_{\rm eq} = \left(\sigma^{(B)}_0  - \frac{e^3}{2\pi^2\hbar^2c}b_0 \right)\mathbf{B}_0 = \frac{e^2 \left(\mu_{5,0} -eb_0 \right)\mathbf{B}_0}{2\pi^2\hbar^2c}= \mathbf{0}.
\end{equation}
This implies that $\mu_{5,0}=eb_0$, i.e. the chiral chemical potential in the global equilibrium state
is unambiguously determined by the energy separation between the Weyl nodes. This also agrees with
the analysis within a band-theoretical model of Weyl semimetals \cite{Franz}.

The inclusion of an external magnetic field does not affect the total electrical neutrality of the sample. This implies, therefore, that the global equilibrium value of the electric chemical potential $\mu_{0,B}$ in the
presence of the magnetic field should be self-consistently determined by the following neutrality condition:
\begin{eqnarray}
\label{Minimal-WH-2-compensation-B}
\frac{\mu_{0,B}\left(\mu_{0,B}^2+3\mu^2_{5,0}+\pi^2T_0^2\right)}{3\pi^2 \hbar^3 v_F^3} +\frac{e^2 (\mathbf{b}\cdot\mathbf{B}_0)}{2\pi^2\hbar^2c^2} -\frac{\mu_{0}\left(\mu^2_{0}+3\mu^2_{5,0}+\pi^2T^2_{0}\right)}{3\pi^2 \hbar^3 v_F^3}=0,
\end{eqnarray}
where we used Eqs.~(\ref{model-rho-def}), (\ref{model-CS-charge}), and the result in Eq.~(\ref{Minimal-WH-2-compensation-B0}).
(Here we assumed that temperature does not change when the magnetic field is included.) The
above equation defines the electric chemical potential $\mu_{0,B}$ as a function of $\mathbf{B}_0$, $\mathbf{b}$,
and $T_0$, as well as the reference value of the electric chemical potential $\mu_0$. Therefore, when
$\mathbf{B}_0 \neq \mathbf{0}$ one should replace $\mu_0$ with $\mu_{0,B}$ in all thermodynamic functions, i.e.
$w_0$, $\epsilon_0$, $P_0$, $n_0$, and $n_{5,0}$.

When a long-wavelength collective mode propagates through the chiral electron fluid, it perturbs the local
values of thermodynamic parameters and fields through the following deviations: $\delta\mu(x)$, $\delta\mu_5(x)$, $\delta T(x)$, and $\delta \mathbf{B}(x)$.
In general, the local fluid velocity $\delta\mathbf{u}(x)$ and the electric field $\delta\mathbf{E}(x)$ are also nonzero in the perturbed state.

As usual in the study of collective excitations, we look for the solutions in the form of
plain waves with the frequency $\omega$ and the wave vector $\mathbf{k}$, i.e. $\delta\mu(x) =
\delta\mu\, e^{-i\omega t+i\mathbf{k}\mathbf{r}}$ and the similar expressions for other perturbed
quantities. To the linear order in perturbations, the electric and chiral charge continuity relations
(\ref{model-J-conserv-eq}) and (\ref{model-J5-conserv-eq}) are given by
\begin{eqnarray}
\label{Minimal-WH-2-conserv-eqs-2-J}
&&\omega \delta \rho-\left(\mathbf{k}\cdot\delta \mathbf{J}\right)=0,\\
\label{Minimal-WH-2-conserv-eqs-2-J5}
&&\omega \delta \rho_{5} -(\mathbf{k}\cdot \delta\mathbf{J}_5)
=- i \frac{e^3}{2\pi^2\hbar^2c} \left(\mathbf{B}_0\cdot\delta\mathbf{E}\right),
\end{eqnarray}
where
\begin{eqnarray}
\label{Minimal-WH-2-J0-def-osc}
\delta \rho &=& -e\delta n +\frac{\left(\mathbf{B}_0\cdot\delta\mathbf{u}\right) \sigma^{(B)}}{3v_F^2}
+i \frac{5 c^2 \sigma^{(\epsilon, u)} (\mathbf{B}_0\cdot [\mathbf{k}\times \delta \mathbf{u}])}{2v_F^2}
-\frac{e^3(\mathbf{b}\cdot\delta\mathbf{B})}{2\pi^2\hbar^2c^2},\\
\label{Minimal-WH-2-J50-def-osc}
\delta \rho_{5} &=& -e\delta n_5 +\frac{\left(\mathbf{B}_0\cdot\delta\mathbf{u}\right) \sigma^{(B)}_5}{3v_F^2} ,
\end{eqnarray}
and
\begin{eqnarray}
\label{Minimal-WH-2-J-def-osc}
\delta\mathbf{J} &=& -en_{0}\delta\mathbf{u} + \mathbf{B}_0\delta\sigma^{(B)}
+ \frac{e^3\left[\mathbf{b}\times\delta\mathbf{E}\right]}{2\pi^2\hbar^2c}
+\frac{i}{2}\sigma^{(V)}\left[\mathbf{k}\times\delta \mathbf{u}\right]
-\frac{1}{4}\sigma^{(\epsilon, V)}\left[\mathbf{k}\times\left[\mathbf{k}\times\delta \mathbf{u}\right]\right],\\
\label{Minimal-WH-2-J5-def-osc}
\delta\mathbf{J}_5 &=& -en_{5,0}\delta\mathbf{u} + \sigma^{(B)}_5 \delta\mathbf{B}
+\mathbf{B}_0\delta\sigma^{(B)}_5
+\frac{i}{2}\sigma^{(V)}_5\left[\mathbf{k}\times\delta \mathbf{u}\right]
-\frac{1}{4}\sigma^{(\epsilon, V)}_5\left[\mathbf{k}\times\left[\mathbf{k}\times\delta \mathbf{u}\right]\right].
\end{eqnarray}
Here we assumed that all coefficients are evaluated for the global equilibrium values
of the thermodynamic parameters, as well as used condition~(\ref{Minimal-WH-2-J0-compensation})
and the expression for the vorticity deviation $\delta \bm{\omega} = i\left[\mathbf{k}\times\delta\mathbf{u}\right]/2$.

The linearized forms of the Euler equation (\ref{model-Euler}) and the energy conservation
relation (\ref{model-energy}) are given by
\begin{eqnarray}
\label{Minimal-WH-2-conserv-eqs-2-Euler-Fourier}
&&\frac{\omega}{v_F}\left\{\frac{ w_0  \delta\mathbf{u}}{v_F}+\mathbf{B}_0\delta\sigma^{(\epsilon,B)} + \sigma^{(\epsilon,B)}\delta\mathbf{B} +i\frac{\hbar n_{5,0}\left[\mathbf{k}\times\delta\mathbf{u}\right]}{4}\right\} -\mathbf{k}\delta P -\frac{4 \sigma^{(\epsilon,B)}}{15v_F}
\left[ \mathbf{k}\left(\mathbf{B}_0\cdot\delta\mathbf{u}\right) +\delta\mathbf{u}\left(\mathbf{B}_0\cdot\mathbf{k}\right) +\mathbf{B}_0\left(\mathbf{k}\cdot\delta\mathbf{u}\right)\right] \nonumber\\
&& -\frac{c \sigma^{(\epsilon, B)}}{3v_F}\left[\mathbf{k}\times\delta\mathbf{E}\right]
-5\sigma^{(\epsilon, u)}\mathbf{k}\left(\mathbf{B}_{0}\cdot\delta \mathbf{B}\right)
-i\frac{\sigma^{(\epsilon, V)}}{10c}
\left\{(\mathbf{B}_0\cdot\mathbf{k})[\mathbf{k}\times\delta\mathbf{u}] +\mathbf{k}\left(\mathbf{B}_0\cdot[\mathbf{k}\times\delta\mathbf{u}]\right)\right\} \nonumber\\
&&=-en_{0}i \delta\mathbf{E} +\frac{i}{c}\left[\mathbf{B}_0\times \left(en_0\delta\mathbf{u}
- \frac{i\sigma^{(V)} [\mathbf{k}\times\delta\mathbf{u}]}{6} \right)\right] -\frac{i w_0  \delta\mathbf{u}}{v_F^2\tau} +\frac{\hbar n_{5,0} [\mathbf{k}\times\delta\mathbf{u}]}{4v_F\tau}
\end{eqnarray}
and
\begin{equation}
\label{Minimal-WH-2-conserv-eqs-2-Energy-Fourier}
\omega \delta\epsilon
- w_0 (\mathbf{k}\cdot\delta\mathbf{u})  -\sigma^{(\epsilon, u)} \left[
\left(\mathbf{B}_0\cdot\delta \mathbf{u}\right)\left(\mathbf{B}_0\cdot \mathbf{k}\right) - 2B_0^2\left(\mathbf{k}\cdot\delta \mathbf{u}\right)\right]
-v_F\left(\mathbf{B}_0\cdot\mathbf{k}\right)\delta\sigma^{(\epsilon,B)} =i\sigma^{(B)}\left(\delta\mathbf{E}\cdot \mathbf{B}_0\right),\\
\end{equation}
respectively. Note that the last equation is not modified by the effects of the vorticity.

Finally, after taking into account Faraday's law $\delta \mathbf{B}=(c/\omega) \left[\mathbf{k}\times \delta \mathbf{E}\right]$,
the remaining Maxwell's equation for $ \delta \mathbf{E}$ is given by
\begin{equation}
\label{Minimal-WH-2-conserv-eqs-Maxwell-1-S}
\left(\omega^2 -\frac{c^2k^2}{ \varepsilon_{e}\mu_m}\right)\delta\mathbf{E} +\mathbf{k}(\mathbf{k}\cdot\delta\mathbf{E})
=\frac{4\pi i \omega}{\varepsilon_{e}} \left(en_{0}\delta\mathbf{u} - \mathbf{B}_0\delta\sigma^{(B)}
- \frac{e^3\left[\mathbf{b}\times\delta\mathbf{E}\right]}{2\pi^2\hbar^2c}
-\frac{i \sigma^{(V)} \left[\mathbf{k}\times\delta\mathbf{u}\right]}{2}
+\frac{\sigma^{(\epsilon, V)}\left[\mathbf{k}\times\left[\mathbf{k}\times\delta \mathbf{u}\right]\right]}{4}
\right).
\end{equation}
In general, the chiral shift $\mathbf{b}$ can be given in components as follows:
$\mathbf{b}=\left(b_{\perp}, \tilde{b}_{\perp}, b_{\parallel}\right)$, where $b_{\perp}$ and $\tilde{b}_{\perp}$
are the two independent components perpendicular to $\mathbf{B}_0$ and $b_{\parallel}$ is
the component parallel to $\mathbf{B}_0$. In the case of the longitudinal (with respect to the
magnetic field) propagation of collective modes, the rotational symmetry can be used to eliminate one of the
perpendicular components and, consequently, we can set $\tilde{b}_{\perp}=0$ without loss
of generality. For transverse modes, however, both $b_{\perp}$ and $\tilde{b}_{\perp}$
are relevant. By definition, we set $b_{\perp}$ to be the component of the chiral shift that
is perpendicular to $\mathbf{B}_0$, but parallel to $\mathbf{k}$, while $\tilde{b}_{\perp}$ is
perpendicular to both $\mathbf{k}$ and $\mathbf{B}_0$.

In the following sections, we will analyze the collective excitations in Dirac and Weyl materials
by using the linearized form of the CHD equations obtained above.

\section{Collective excitations in the absence of external magnetic field}
\label{sec:WMHD-B0}

In this section, we study the collective excitations at $\mathbf{B}_0=\mathbf{0}$. Even in this rather simple case, the spectrum of
collective excitations is affected by the chiral shift and contains qualitatively new modes.

\subsection{Dirac semimetals and PI symmetry broken Weyl semimetals}
\label{sec:WMHD-B0-b0}

In order to test the CHD, let us start from the simplest case of Dirac semimetals.
Therefore, we set $|\mathbf{b}|=b_0=0$, as well as $\mu_{5,0}=0$ in accordance with
Eq.~(\ref{Minimal-WH-2-J0-compensation}). We also note that, according to
Eq.~(\ref{Minimal-WH-2-compensation-B}), we have $\mu_{0,B}=\mu_0$ in the absence of a background magnetic field.

When the electric and chiral charge densities vanish (i.e. at $\mu_0=\mu_{5,0}=0$),
we obtain the following roots of the characteristic equation [which is the determinant of the system (\ref{Minimal-WH-2-conserv-eqs-2-J}), (\ref{Minimal-WH-2-conserv-eqs-2-J5}), and (\ref{Minimal-WH-2-conserv-eqs-2-Euler-Fourier})--(\ref{Minimal-WH-2-conserv-eqs-Maxwell-1-S})]:
\begin{eqnarray}
\label{WMHD-B0-b0-mu0mu50-omega-d}
\omega_{\rm d} &=& -\frac{i}{\tau},\\
\label{WMHD-B0-b0-mu0mu50-omega-s}
\omega_{\rm s, \pm} &=& -\frac{i}{2\tau} \pm i\frac{\sqrt{3-4\tau^2v_F^2|\mathbf{k}|^2}}{2\sqrt{3}\tau},
\end{eqnarray}
and the roots corresponding to the standard in-medium electromagnetic (light) waves, i.e. $\omega_{\rm light, \pm} =\pm c|\mathbf{k}|/\sqrt{\varepsilon_e \mu_m}$. Here $\varepsilon_e $ is the electric permittivity and $\mu_m$ is the magnetic permeability. As is easy to check, the latter roots are associated with oscillations
of electromagnetic fields, but induce no hydrodynamic motion of the electron fluid. In a strict sense,
therefore, they are not hydrodynamic modes. As we will see later, however, under certain conditions
the electromagnetic waves could hybridize with the electron fluid oscillations and produce new types of collective
excitations.

The first solution, $\omega_{\rm d}$, corresponds to a doubly degenerate diffusive mode. The next
pair of solutions in Eq.~(\ref{WMHD-B0-b0-mu0mu50-omega-s}) describes damped sound waves.
As is easy to check, in the limit of large $\tau$, the corresponding frequencies $\omega_{\rm s, \pm}$
approach the standard dispersion relation for sound waves in plasma, $\omega_{\rm s, \pm} \simeq
\pm v_F|\mathbf{k}|/\sqrt{3}$. By analyzing the eigenstates of these sound waves, one can verify
that, as expected, they are sustained by oscillations of the fluid velocity and do not induce any local
electromagnetic field.

The spectrum changes in the case of a nonzero electric charge density $\rho_{0}$ (i.e. at $\mu_0\neq 0$,
but still $\mu_{5,0}=0$). In the long-wavelength limit ($|\mathbf{k}|\to 0 $), the corresponding dispersion relations
of collective modes in Dirac semimetals read
\begin{eqnarray}
\label{WMHD-B0-b0-mu50-omega-d}
\omega_{\rm d, \pm}&\approx&
-\frac{i}{2\tau} \pm \frac{i}{2\tau}\sqrt{1-\frac{16\pi v_F^2 \rho_{0}^2 \tau^2}{\varepsilon_e  w_0 }}
+O(|\mathbf{k}|),\\
\label{WMHD-B0-b0-mu50-omega-s}
\omega_{\rm s, \pm} &=& -\frac{i}{2\tau} \pm \frac{i}{2\tau}\sqrt{1-\frac{4\tau^2 v_F^2 \left(12\pi \rho_{0}^2 +w_0|\mathbf{k}|^2\right)}{3\varepsilon_e  w_0 }},\\
\label{WMHD-B0-b0-mu50-omega-d12}
\omega_{\rm d,1} &=&\omega_{\rm d,2} \approx -i\frac{c^2|\mathbf{k}|^2  w_0 }{4\pi \mu_m v_F^2 \rho_{0}^2 \tau }+O(|\mathbf{k}|^3).
\end{eqnarray}
First, as is clear from the expressions for $\omega_{\rm d, \pm}$ and $\omega_{\rm s, \pm}$ in Eqs.~(\ref{WMHD-B0-b0-mu50-omega-d}) and (\ref{WMHD-B0-b0-mu50-omega-s}),
a nonzero electric charge density makes the dissipative and sound waves gapped. [Note that each of the former modes is doubly degenerate.]
In addition to these gapped modes, there are also gapless
doubly degenerate diffusive waves with $\omega = \omega_{\rm d,1}$ and $\omega = \omega_{\rm d,2}$,
which appear to be hybridized electromagnetic waves.

The results here can be easily generalized to the case of Weyl semimetals with a broken PI (i.e. with $b_0\neq 0$),
but intact TR  symmetry (i.e. $|\mathbf{b}|=0$).
In such a case, in accordance with Eq.~(\ref{Minimal-WH-2-J0-compensation}),
the chiral chemical potential is nonzero in equilibrium, i.e. $\mu_{5,0}=eb_0\neq0$. When the electric charge density
is also nonzero, we find that the frequencies of the collective modes in Eqs.~(\ref{WMHD-B0-b0-mu50-omega-d}),
(\ref{WMHD-B0-b0-mu50-omega-s}), and (\ref{WMHD-B0-b0-mu50-omega-d12}) remain unchanged to the leading
order in the wave vector. However, in the limit of vanishing electric charge density (i.e. $\mu_0=0$, but $\mu_{5,0}\neq 0$),
the long-wavelength spectrum of the diffusive waves is modified by a nonzero $n_{5,0}$. In particular,
the corresponding frequencies are given by
\begin{equation}
\label{WMHD-B0-b0-mu0-omega-d}
\omega_{\rm d,\pm} = -\frac{i}{\tau} \frac{4 w_0  \mp \hbar n_{5,0} v_F |\mathbf{k}|}{4 w_0
\pm \hbar n_{5,0} v_F |\mathbf{k}|}.
\end{equation}
As for the sound and light waves, they remain the same, see Eq.~(\ref{WMHD-B0-b0-mu0mu50-omega-s}) and the text after it.

\subsection{Weyl semimetals with broken TR symmetry}
\label{sec:WMHD-B0-all-b}

In this subsection, we consider the case of Weyl semimetals with a broken TR symmetry, which is characterized
by $|\mathbf{b}|\neq0$. In the simplest case of vanishing chemical potentials $\mu_0=\mu_{5,0}=0$, we find
that the frequencies of both diffusive and sound waves coincide with those in a Dirac semimetal, see
Eqs.~(\ref{WMHD-B0-b0-mu0mu50-omega-d}) and (\ref{WMHD-B0-b0-mu0mu50-omega-s}).
In addition, there are also collective modes that strongly depend on the chiral shift. Their dispersion relations read
\begin{eqnarray}
\label{WMHD-B0-all-b-mu0mu50-omega-AHW-0}
\omega_{\text{{\tiny AHW}}, \pm} &=& \pm \frac{\sqrt{2e^6\mu_m |\mathbf{b}|^2 + \pi^2 c^4\varepsilon_e \hbar^4 |\mathbf{k}|^2 - 2e^3\sqrt{\mu_m\left(e^6 \mu_m |\mathbf{b}|^4 +\pi^2 c^4\varepsilon_e \hbar^4 \left(\mathbf{k}\cdot\mathbf{b}\right)^2\right)}}}{\pi c \varepsilon_e \hbar^2 \sqrt{\mu_m}},\\
\label{WMHD-B0-all-b-mu0mu50-omega-gAHW}
\omega_{\text{{\tiny gAHW}}, \pm} &=& \pm \frac{\sqrt{2e^6\mu_m |\mathbf{b}|^2 + \pi^2 c^4\varepsilon_e \hbar^4 |\mathbf{k}|^2 + 2e^3\sqrt{\mu_m \left(e^6 \mu_m |\mathbf{b}|^4 +\pi^2 c^4\varepsilon_e \hbar^4 \left(\mathbf{k}\cdot\mathbf{b}\right)^2\right)}}}{\pi c \varepsilon_e \hbar^2 \sqrt{\mu_m}}.
\end{eqnarray}
These are the electromagnetic (light) modes modified by the Chern--Simons contributions in Maxwell's
equations. (In the limit $|\mathbf{b}|\to0$, both modes reduce to the usual in-medium light waves.) As is
easy to check, the propagation of these modes is affected by a nonzero chiral shift
$\mathbf{b}$ via the anomalous Hall effect currents $\propto [\mathbf{b}\times \mathbf{E}]$. Such currents
mix the longitudinal and transverse (with respect to the direction of the wave vector) components of the
oscillating electromagnetic fields. While one of the modes, namely the \emph{anomalous Hall wave} (AHW) with the
frequency $\omega_{\text{{\tiny AHW}}, \pm}$ given by Eq.~(\ref{WMHD-B0-all-b-mu0mu50-omega-AHW-0}), remains gappless, the other one,
namely the \emph{gapped anomalous Hall wave} (gAHW) with the frequency $\omega_{\text{{\tiny gAHW}}, \pm}$ given by Eq.~(\ref{WMHD-B0-all-b-mu0mu50-omega-gAHW})
acquires a gap $2e^3|\mathbf{b}|/\left(\pi c \varepsilon_e \hbar^2\right)$.
As expected, both types of the anomalous Hall waves
are nondissipative. Also, strictly speaking, these waves are nonhydrodynamic modes since they are not accompanied by fluid oscillations.

It should be pointed that the dispersion relations of both anomalous Hall waves depend on the direction of their
propagation with respect to the orientation of the chiral shift. This dependence is particularly strong in
the case of the gapless mode. Indeed, when the AHW propagates perpendicularly to $\mathbf{b}$
(i.e. $\mathbf{k}\perp \mathbf{b}$), it reduces to the regular in-medium light wave, i.e.
\begin{equation}
\label{WMHD-B0-all-b-mu0mu50-omega-AHW-light}
\lim_{k_{\parallel}\to0}\omega_{\text{{\tiny AHW}}, \pm} = \omega_{\rm light, \pm}.
\end{equation}
However, for $\mathbf{k}\parallel \mathbf{b}$, the AHW
turns into the \emph{anomalous helicon} (AH) with a quadratic dispersion relation, i.e.
\begin{equation}
\label{WMHD-B0-all-b-mu0mu50-omega-AHW}
\lim_{k_{\perp}\to0}\omega_{\text{{\tiny AHW}}, \pm} = \omega_{\text{{\tiny AH}}, \pm} \approx \pm \frac{\pi c^3\hbar^2 k_{\parallel}^2}{2e^3\mu_m |\mathbf{b}|} +O(|\mathbf{k}|^3).
\end{equation}
It should be emphasized that the limit $ |\mathbf{b}|\to 0 $ cannot be taken in the last expression,
which is valid only at long wavelengths. However, such a limit is well defined in the exact
expression (\ref{WMHD-B0-all-b-mu0mu50-omega-AHW-0}) and, as expected, reproduces the dispersion
relation of the usual light wave.

The result in Eq.~(\ref{WMHD-B0-all-b-mu0mu50-omega-AHW}) is quite remarkable. It shows that the helicon-type
excitations can exist in Weyl semimetals without any background magnetic field and at vanishing electric charge
density. This is in drastic contrast to usual plasmas, in which helicons exist only at nonzero magnetic field and electric
charge density. In connection to this AH mode, we should remark that a similar dispersion relation can be formally
obtained by taking the limit of a vanishing background magnetic field in the result found in Ref.~\cite{Pellegrino}, where the CKT approach was used. However,
we note that the corresponding conclusion is not reliable
because the analysis in Ref.~\cite{Pellegrino}
heavily relies on the presence of a cyclotron resonance and, consequently, the presence of a magnetic field.

It is also interesting to consider the case of Weyl semimetals with nonzero electric charge density (i.e. $\mu_0\neq0$).
By setting $\mu_{5,0}=0$, we find the following frequencies of the diffusive wave and the gAHW in the long-wavelength limit:
\begin{eqnarray}
\label{WMHD-B0-all-b-mu50-omega-d}
\omega_{\rm d, \pm} &\approx& -\frac{i}{2\tau} \pm \frac{e^3 |\mathbf{b}|}{c\pi \varepsilon_e \hbar^2} - \frac{i \sqrt{ w_0 \left(\pi c \varepsilon_e \hbar^2 \mp 2i\tau e^3 |\mathbf{b}| \right)^2 -16\pi^3\varepsilon_e c^2v_F^2 \hbar^4 \rho_{0}^2 \tau^2}}{2\pi c \varepsilon_e \hbar^2 \tau \sqrt{ w_0 }}
+O(|\mathbf{k}|),\\
\label{WMHD-B0-all-b-mu50-omega-gAHW}
\omega_{\text{{\tiny gAHW}}, \pm} &\approx& -\frac{i}{2\tau} \pm \frac{e^3 |\mathbf{b}|}{c\pi \varepsilon_e \hbar^2} + \frac{i \sqrt{ w_0 \left(\pi c \varepsilon_e \hbar^2 \mp 2i\tau e^3 |\mathbf{b}| \right)^2 -16\pi^3\varepsilon_e c^2v_F^2 \hbar^4 \rho_{0}^2 \tau^2}}{2\pi c \varepsilon_e \hbar^2 \tau \sqrt{ w_0 }} +O(|\mathbf{k}|).
\end{eqnarray}
In the same limit, the frequencies of the AHW are given by Eq.~(\ref{WMHD-B0-all-b-mu50-omega-AHW}) in Appendix~\ref{sec:app-formulas-frequencies} [we will discuss a few limiting cases below].
In addition, while the gap in the sound frequencies $\omega_{\rm s, \pm}$ is still given by Eq.~(\ref{WMHD-B0-b0-mu50-omega-s}), the dependency on the wave vector is different.
Note that all these modes are the generalized versions of those at zero charge density.

In order to clarify the nature of the AHW, it is instructive to consider the two limiting cases: $\mathbf{k}\perp \mathbf{b}$ ($b_{\parallel}\to0$) and $\mathbf{k} \parallel \mathbf{b}$ ($b_{\perp}\to0$).
In the limit $b_{\parallel}\to0$, the AHW becomes completely
diffusive with the following frequencies:
\begin{eqnarray}
\label{WMHD-B0-bx-mu50-omega-d1}
\lim_{b_{\parallel}\to0}\omega_{\text{{\tiny AHW}}, +} &=&\omega_{\rm d,1} \approx
-i\frac{c^2|\mathbf{k}|^2  w_0 }{4\pi \mu_m v_F^2 \rho_{0}^2 \tau}+O(|\mathbf{k}|^3),\\
\label{WMHD-B0-bx-mu50-omega-d2}
\lim_{b_{\parallel}\to0}\omega_{\text{{\tiny AHW}}, -} &=&\omega_{\rm d,2} \approx
-i\frac{v_F^2|\mathbf{k}|^2  w_0  \tau \left(3\pi^3 c^4 \hbar^4 \rho_0^2 +e^6\mu_m w_0  b_{\perp}^2 \right)}{3\mu_m \left(4\pi^4 c^2v_F^4 \hbar^4 \rho_0^4\tau^2 +e^6  w_0 ^2 b_{\perp}^2\right)}+O(|\mathbf{k}|^3).
\end{eqnarray}
The degeneracy of these diffusive waves is lifted by a nonzero component of the chiral shift $b_{\perp}$.

The same modes in the limit $b_{\perp}\to0$ are qualitatively different. In this case, instead of the completely diffusive waves,
we obtain the AH modes with frequencies
\begin{equation}
\label{WMHD-B0-bz-mu50-omega-h}
\lim_{b_{\perp}\to0}\omega_{\text{{\tiny AHW}}, \pm}=\omega_{\text{{\tiny AH}}, \pm} \approx
- \frac{\pi c^3 \hbar^2  w_0  |\mathbf{k}|^2 \left(2 i\pi^2 c  \hbar^2 v_F^2 \rho_0^2 \tau \mp w_0 e^3b_{\parallel} \right)}{2 \mu_m\left(4\pi^4 c^2 v_F^4\hbar^4 \rho_0^4 \tau^2 +e^6  w_0 ^2 b_{\parallel}^2 \right)}  + O(|\mathbf{k}|^3).
\end{equation}
By comparing the above result with that in Eq.~(\ref{WMHD-B0-all-b-mu0mu50-omega-AHW}), we find that the AH
becomes dissipative, with the imaginary part determined by the electric charge density.

For completeness, let us also briefly mention what happens in the case of Weyl semimetals where both TR and PI
symmetries are broken. Such Weyl semimetals are characterized not only by $|\mathbf{b}|\neq 0$, but also a nonzero
chiral chemical potential $\mu_{5,0}=eb_0\neq0$. At nonzero electric charge density, we find that the expressions
for $\omega_{\text{{\tiny AHW}}, \pm}$, $\omega_{\rm d, \pm}$, and $\omega_{\text{{\tiny gAHW}}, \pm}$  [see Eqs.~(\ref{WMHD-B0-all-b-mu50-omega-AHW}), (\ref{WMHD-B0-all-b-mu50-omega-d}), and (\ref{WMHD-B0-all-b-mu50-omega-gAHW})] remain unchanged to the leading order in the wave vector.
As for the case of the vanishing electric charge density
(i.e. $\mu_0=0$), the only mode affected by a nonzero $\mu_{5,0}$ will be the diffusive wave given in
Eq.~(\ref{WMHD-B0-b0-mu0-omega-d}).

\begin{figure}[t]
\begin{center}
\hspace{-0.45\textwidth}(a)\hspace{0.45\textwidth}(b)\\[0pt]
\includegraphics[width=0.45\textwidth]{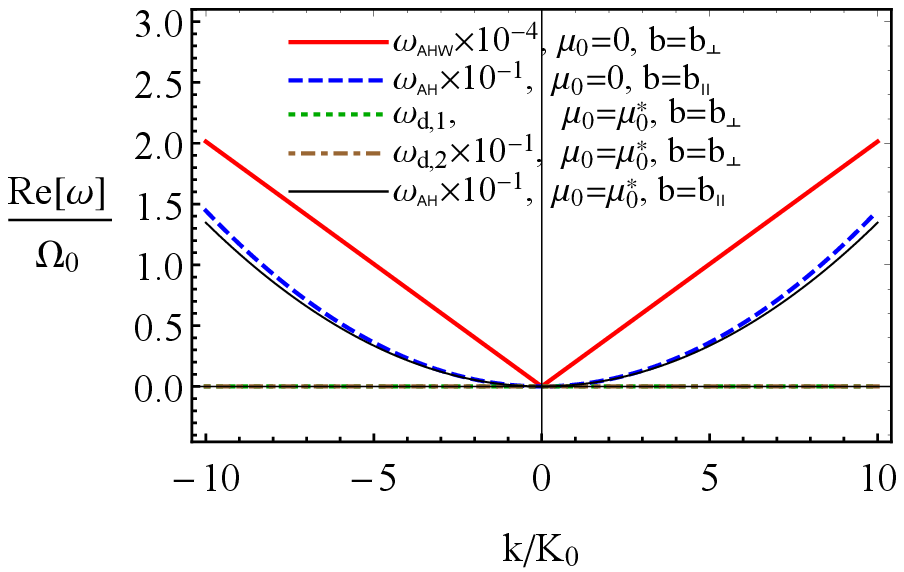}
\hfill
\includegraphics[width=0.45\textwidth]{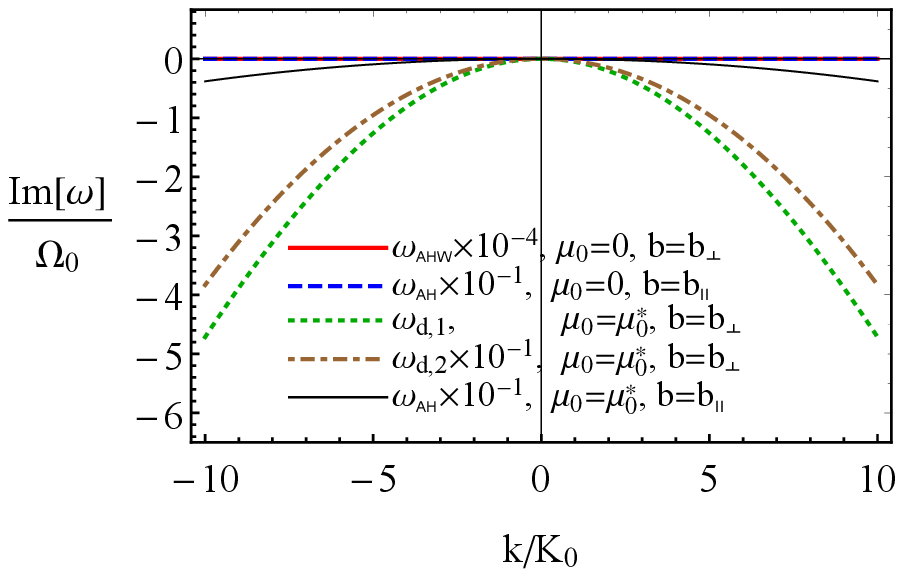}
\caption{The real (panel a) and imaginary (panel b) parts of the AHW and AH frequencies.
Red solid lines correspond to the case $\mathbf{b}\perp\mathbf{k}$ and $\mu_{0}=0$,
blue dashed lines represent $\omega_{\text{{\tiny AH,+}}}$ at $\mathbf{b}\parallel\mathbf{k}$
and $\mu_{0}=0$, green dotted and brown dot-dashed lines correspond to
$\omega_{\text{{\tiny AHW}},+}=\omega_{d,1}$ and $\omega_{\text{{\tiny AHW}},-}=\omega_{d,2}$,
respectively, at $\mathbf{b}\perp\mathbf{k}$ and $\mu_{0}=\mu_{0}^{*}$, and
the solid black lines show $\omega_{\text{{\tiny AH}}}$ at $\mathbf{b}\parallel\mathbf{k}$ and $\mu_{0}=\mu_{0}^{*}$.
We used the model parameters in Eq.~(\ref{App-realistic-parameters}),
as well as $\mu_{0}^{*}=10~\mbox{meV}$, $\mu_{5,0}=0~\mbox{meV}$, $T_0=10~\mbox{K}$, and $B_0=0$.
The values of the wave vectors and the frequencies are given in units of $K_{0}=10^{-6}\pi/a \approx12.3~\mbox{cm}^{-1}$
and $\Omega_{0}=v_F K_0\approx1.84~\mbox{GHz}$.}
\label{fig:WMHD-B0-AHW}
\end{center}
\end{figure}

The real and imaginary parts of the collective modes frequencies for several choices of $\mathbf{b}$
and $\mu_0$ are shown in Figs.~\ref{fig:WMHD-B0-AHW}(a) and \ref{fig:WMHD-B0-AHW}(b), respectively.
For our numerical estimates, we set the electric permittivity $\varepsilon_e=1$ and the magnetic permeability $\mu_m = 1$, as well as used the following representative set of model parameters:
\begin{equation}
\label{App-realistic-parameters}
v_F\approx1.49\times10^8~\mbox{cm/s}, \qquad
\tau =10^{-12}~\mbox{s}, \qquad a \approx 25.5\times10^{-8}~\mbox{cm}, \qquad
b= \frac{\hbar c}{e} b_{\rm latt} , \qquad
b_{\rm latt} \approx 0.3\frac{\pi}{a},
\end{equation}
which are close to the parameters that describe the low-energy spectrum of electron quasiparticles
in Cd$_3$As$_2$ \cite{Neupane:2014,Zhang-Xiu:2015}. Note also that, in general, the relaxation time is a function of chemical potentials and temperature. However, henceforth, we will neglect such a dependence in order to present our results as clearly as possible.

The values of the wave vectors and frequencies in Figs.~\ref{fig:WMHD-B0-AHW}(a) and \ref{fig:WMHD-B0-AHW}(b) are given in units of
$K_{0}=10^{-6}\pi/a \approx12.3~\mbox{cm}^{-1}$ and $\Omega_{0}=v_F K_0\approx1.84~\mbox{GHz}$,
which can be viewed as the characteristic values in the problem at hand. In order to avoid the clutter,
we did not show the dispersion relation of the gAHW in the figure. Numerically, the corresponding mode
is weakly dispersing with $\mbox{Re}\left(\omega_{\text{{\tiny gAHW}}}\right)\approx 2e^3|\mathbf{b}|/(\pi c \varepsilon_e \hbar^2)$ and has a relatively small imaginary part at $\mu_0\neq0$.

Let us briefly discuss the dispersion relations of the collective modes presented in Figs.~\ref{fig:WMHD-B0-AHW}(a) and \ref{fig:WMHD-B0-AHW}(b).
At $\mu_{0}=0$, the AHWs are dissipationless and have linear dispersion laws when
$\mathbf{k}\perp\mathbf{b}$ and quadratic ones when $\mathbf{k}\parallel\mathbf{b}$. While
the former case describes nothing else but the usual in-medium light waves, the latter is the AH mode given by Eq.~(\ref{WMHD-B0-all-b-mu0mu50-omega-AHW}), whose existence
is a distinctive feature of Weyl semimetals with a broken TR symmetry. When
$\mu_0\neq0$ and $b=b_{\perp}$, the AHWs are completely diffusive gapless waves, with
frequencies $\omega_{\text{{\tiny AHW}},+}=\omega_{d,1}$ and $\omega_{\text{{\tiny AHW}},-}=\omega_{d,2}$
given by Eqs.~(\ref{WMHD-B0-bx-mu50-omega-d1}) and (\ref{WMHD-B0-bx-mu50-omega-d2}), respectively.
In addition, while the real part of the AH frequency given by Eq.~(\ref{WMHD-B0-bz-mu50-omega-h})
is almost unaffected by a nonzero $\mu_0$, a small imaginary part develops and makes the wave slightly
damped.
Last but not least, we checked that a nonzero $\mu_{5,0}$, even as large as $\mu_{0}$, leads to rather small quantitative changes in the spectrum of the collective excitations.

\section{Longitudinal propagation of collective modes}
\label{sec:WMHD-B-kz}

In this section, we consider the case of the longitudinal propagation of collective modes with respect to
the external magnetic field $\mathbf{B}_0$ (i.e. $\mathbf{k}\parallel \mathbf{B}_0$).

\subsection{Dirac semimetals and PI symmetry broken Weyl semimetals}
\label{sec:WMHD-B-kz-b0}

It is instructive to start our analysis from the
simplest case of Dirac semimetals, i.e. $|\mathbf{b}|=b_0=0$. At $\mu_0=\mu_{5,0}=0$, the spectrum of the longitudinal collective modes can be obtained analytically.
We find, in particular, that the frequencies of the diffusive and electromagnetic
(light) waves are unaffected by $\mathbf{B}_0$, see Eq.~(\ref{WMHD-B0-b0-mu0mu50-omega-d}) and the text after it.
On the other hand, the dispersion relation of the
sound wave changes and is now given by
\begin{equation}
\label{WMHD-B-kx-b0-mu0mu50-omega-s-general-k}
\omega_{\rm s, \pm} = -\frac{i}{2\tau} \pm \frac{i}{2\tau} \sqrt{1 -4\tau^2 v_F^2\frac{|\mathbf{k}|^2 w_0
-\sigma^{(\epsilon, u)} \left[2|\mathbf{k}|^2B_0^2 -(\mathbf{k}\cdot\mathbf{B}_0)\right]}{3 w_0 }},
\end{equation}
which is applicable for both longitudinal and transverse propagation of the sound waves.
In addition, there is the \emph{gapped chiral magnetic wave} (gCMW) with the frequency
\begin{equation}
\label{WMHD-B-kz-b0-mu0mu50-omega-gCMW}
\omega_{\text{{\tiny gCMW}},\pm} = \pm \frac{eB_0 \sqrt{3v_F^3 \left(4\pi e^2T_0^2 +3\varepsilon_e \hbar^3 v_F^3 k_{\parallel}^2\right)}}{2\pi^2T_0^2 c \sqrt{\varepsilon_e \hbar}}.
\end{equation}
The gap of this excitation is not identical to, but shares some similarity with the cyclotron frequency $\Omega_c\sim B_0/T_0$,
which is also proportional to the magnetic field and inversely proportional to temperature. While, naively, this mode can be
identified with the cyclotron wave in the chiral electron plasma, it could be also related to the chiral magnetic wave (CMW) \cite{Kharzeev}. Indeed, the gCMW is connected with the coupled oscillations of the electric and chiral chemical potentials amended by the oscillating electric field along $\mathbf{B}_0$. However, unlike the conventional CMW, the gCMW includes the effects of the dynamical electromagnetism.

The dispersion relations of the collective modes become rather complicated at nonzero electric charge density,
which is quantified by $\mu_0\neq0$. In total, we found 8 nontrivial solutions of the characteristic equation.
Here, for the sake of clarity and brevity, we analyze only the most interesting of them, i.e.
\begin{eqnarray}
\label{WMHD-B-kz-b0-mu50-omega-d}
\omega_{\rm d,\pm} &\approx& -\frac{i}{2\tau} \mp \frac{v_F^2 \rho_{0} B_0}{2c w_0} - \frac{i \sqrt{\varepsilon_e \left[c  w_0 \mp  i\tau v_F^2 \rho_0 B_0\right]^2 -16\pi v_F^2c^2\rho_{0}^2  w_0 \tau^2 }}{2c\sqrt{\varepsilon_e} w_0 \tau} +O(k_{\parallel}),\\
\label{WMHD-B-kz-b0-mu50-omega-gCMW}
\omega_{\text{{\tiny gCMW}}, \pm} &\approx& -\frac{i}{2\tau} \mp \frac{v_F^2 \rho_{0} B_0}{2c w_0} + \frac{i \sqrt{\varepsilon_e \left[c w_0 \mp  i\tau v_F^2 \rho_0 B_0\right]^2 -16\pi v_F^2c^2\rho_{0}^2 w_0 \tau^2 }}{2c\sqrt{\varepsilon_e} w_0 \tau} +O(k_{\parallel}),\\
\label{WMHD-B-kz-b0-mu50-omega-h}
\omega_{\rm h, \pm} &\approx& \mp  \frac{ck_{\parallel}^2B_0}{4\pi \mu_m \rho_0} -\frac{i}{\tau}
\frac{c^2k_{\parallel}^2 w_0 }{4\pi\mu_m v_F^2 \rho_{0}^2} +O(k_{\parallel}^3).
\end{eqnarray}
By comparing the dispersion relations $\omega_{\rm gAHW,\pm}$ in Eq.~(\ref{WMHD-B0-all-b-mu50-omega-gAHW}) and $\omega_{\text{{\tiny gCMW}}, \pm}$ in Eq.~(\ref{WMHD-B-kz-b0-mu50-omega-gCMW}),
we see that the chiral shift at $B_0=0$ in the former
plays a similar role to that of the magnetic field $\mathbf{B}_0$ at $|\mathbf{b}|=0$ in the latter.
This indicates that the gAHW and gCMW are closely related collective modes.

These results can be easily generalized to the case of Weyl semimetals with a broken PI symmetry. This can be done
by including a nonzero chiral chemical potential, i.e. $\mu_{5,0}=eb_0\neq0$. As is easy to check, in
this case, the diffusive and sound waves are modified. The frequency of the former is given in
Eq.~(\ref{WMHD-B0-b0-mu0-omega-d}). The dispersion relations of the sound waves, on the other
hand, are
\begin{eqnarray}
\label{WMHD-B-kz-b0-mu0-omega-s-p}
\omega_{\rm s, +} &\approx& -i\frac{v_F^2k_{\parallel}^2 \tau (15\mu_{5,0}^2 +7\pi^2T_0^2) \left( w_0 -\mu_{5,0}n_{5,0}-\sigma^{(\epsilon, u)}B_0^2\right)}{3 w_0
(5\mu_{5,0}^2 +7\pi^2T_0^2)
} +O(k_{\parallel}^3),\\
\label{WMHD-B-kz-b0-mu0-omega-s-m}
\omega_{\rm s, -} &\approx& -i\frac{2c w_0 \left(3\mu_{5,0}^2+\pi^2T_0^2\right)}{\tau\left[2c w_0 \left(3\mu_{5,0}^2+\pi^2T_0^2\right) -v_F \hbar \mu_{5,0}\sigma^{(B)} B_0^2 \right]} +O(k_{\parallel}).
\end{eqnarray}
In addition to the conventional electromagnetic (light) wave, there is also the gCMW with the frequency
\begin{equation}
\label{WMHD-B-kz-b0-mu0-omega-gCMW}
\omega_{\text{{\tiny gCMW}},\pm} = \pm \frac{e^2B_0\sqrt{5\mu_{5,0}^2+7\pi^2T^2_0}}{c\pi \hbar^2\sqrt{\varepsilon_e g_0}} +O(k_{\parallel}),
\end{equation}
where used the following shorthand notation:
\begin{equation}
\label{WMHD-B-Q-def}
g_0= \frac{15\mu_{5,0}^4+6\pi^2\mu_{5,0}^2T_0^2+7\pi^4T_0^4}{3\pi v_F^3 \hbar^3}.
\end{equation}
In general, however, we find that all modifications in the spectra of collective modes are quantitative,
rather than qualitative.

When both electric and chiral charge densities are nonzero, we checked that the frequencies
$\omega_{\rm d, \pm}$, $\omega_{\text{{\tiny gCMW}},\pm}$, and $\omega_{\rm h, \pm}$ remain unchanged to the leading order in
small $|\mathbf{k}|$. For their explicit expressions, see  Eqs.~(\ref{WMHD-B-kz-b0-mu50-omega-d}), (\ref{WMHD-B-kz-b0-mu50-omega-gCMW}), and (\ref{WMHD-B-kz-b0-mu50-omega-h}), respectively.

\subsection{Weyl semimetals with $\mathbf{b}\perp\mathbf{B}_0$}
\label{sec:WMHD-B-kz-bx}

Let us now consider the case of the longitudinal waves (i.e. $\mathbf{k}\parallel \mathbf{B}_0$) in
Weyl semimetals with $\mathbf{b}\perp\mathbf{B}_0$ and $|\mathbf{b}|=b_{\perp}$.
The analysis is relatively simple for zero electric and chiral charge densities.
We find that the frequencies of the
diffusive, sound, and light modes are given by the same expressions as in the case of Dirac semimetals,
see Eqs.~(\ref{WMHD-B0-b0-mu0mu50-omega-d}), (\ref{WMHD-B-kx-b0-mu0mu50-omega-s-general-k}), as well as
the text after Eq.~(\ref{WMHD-B0-b0-mu0mu50-omega-d}).
On the other hand, the chiral shift changes
the dispersion relations of the gCMW by turning it into the gAHW with the following frequency:
\begin{equation}
\label{WMHD-B-kz-bx-mu0mu50-omega-gCMW}
\omega_{\text{{\tiny gAHW}},\pm} \approx \pm \frac{e^2 \sqrt{3\varepsilon_e v_F^3 \hbar^3 B_0^2 +4\pi e^2 T^2_0 b_{\perp}^2}}{c\varepsilon_e \hbar^2 T_0 \sqrt{\pi^3}}
+O(k_{\parallel}^2).
\end{equation}
The corresponding exact expression is given by Eq.~(\ref{WMHD-B-kz-bx-mu0mu50-omega-gCMW-app}) in Appendix~\ref{sec:app-formulas}.
We note that both $B_0$ and $b_{\perp}$ play a similar role in producing a nonzero gap for the electromagnetic (light)
mode and, thus, transforming it into the gAHW.

Interestingly, $b_{\perp}$ also gives rise to a qualitatively different type of the AHW, which we call the
\emph{longitudinal anomalous Hall wave} (lAHW).
(The transverse AHW, which was predicted
in Ref.~\cite{Gorbar:2017vph}, will be analyzed in Sec.~\ref{sec:WMHD-B-kx-all-b-mu0mu50}.) While the exact frequency is given by Eq.~(\ref{WMHD-B-kz-bx-mu0mu50-omega-AHW-app}) in Appendix~\ref{sec:app-formulas}, the corresponding result in the long-wavelength limit reads
\begin{equation}
\label{WMHD-B-kz-bx-mu0mu50-omega-AHW}
\omega_{\text{{\tiny lAHW}},\pm} \approx \pm \frac{\hbar B_0 k_{\parallel} \sqrt{3v_F^3 \left(\pi^3 c^4 \hbar T^2_0 +3e^4\mu_m v_F^3 b_{\perp}^2\right)}}{cT_0\sqrt{\pi^3\mu_m \left(3\varepsilon_e v_F^3 \hbar^3 B_0^2 +4\pi e^2T^2_0b_{\perp}^2 \right)}} +O(k_{\parallel}^3).
\end{equation}
Some features of the lAHW resemble those of a general AHW, but the roles that the chiral shift component $b_{\perp}$ and the magnetic
field strength $B_0$ play are different. Furthermore, unlike the AHW, the lAHW has the characteristic linear dispersion law.

In order to clarify the origin of the lAHW, let us present the corresponding nontrivial equations
that describe this mode within the CHD. In the case of $\mu_0=\mu_{5,0}=0$
and $\mathbf{b}\perp\mathbf{B}_0$, they read
\begin{eqnarray}
\label{WMHD-B-kz-bx-mu0mu50-AHW-system-be}
&& \frac{T^2\omega}{3 v_F^3 \hbar} \delta\mu+  \frac{eB_0 k_{\parallel}}{2\pi^2c} \delta\mu_5 =0,\\
\label{WMHD-B-kz-bx-mu0mu50-AHW-system-2}
&&\frac{eB_0 k_{\parallel}}{2\pi^2 c} \delta\mu + \frac{T^2 \omega}{3v_F^3\hbar}\delta\mu_5 -i \frac{e^2B_0}{2\pi^2c} \delta E_{\parallel} =0,\\
\label{WMHD-B-kz-bx-mu0mu50-AHW-system-3}
&&\left(\omega^2  -\frac{c^2k^2_{\parallel}}{\varepsilon_e \mu_m}\right)\delta \tilde{E}_{\perp} - i\frac{2e^3\omega b_{\perp}}{\pi c\varepsilon_e \hbar^2}\delta E_{\parallel}=0,\\
\label{WMHD-B-kz-bx-mu0mu50-AHW-system-ee}
&&\varepsilon_e\omega \delta E_{\parallel} + i\frac{2e^2}{\pi c \hbar^2}\left(B_0 \delta \mu_5 +eb_{\perp}\delta \tilde{E}_{\perp}\right)=0,
\end{eqnarray}
where $\delta \tilde{E}_{\perp}$ denotes the component of the oscillating electric field parallel to $\left[\mathbf{B}_0\times\mathbf{b}\right]$.
The first two equations are the continuity relations for the electric and chiral charges, respectively. The other two
come from Maxwell's equations. We also used Faraday's law $\delta \mathbf{B}=(c/\omega) [\mathbf{k}
\times \delta \mathbf{E}]$, as well as took into account that $\delta T=0$ and $\delta \mathbf{u}=\mathbf{0}$
for this particular AHW mode.

From the continuity equations (\ref{WMHD-B-kz-bx-mu0mu50-AHW-system-be}) and (\ref{WMHD-B-kz-bx-mu0mu50-AHW-system-2}),
we see that the oscillations of the electric and chiral chemical potentials are coupled to each other.
However, because of the chiral anomaly, they also drive the oscillations of $\delta E_{\parallel}$.
Then, due to the AHE, represented by the terms proportional to $b_{\perp}$ in Maxwell's equations (\ref{WMHD-B-kz-bx-mu0mu50-AHW-system-3}) and (\ref{WMHD-B-kz-bx-mu0mu50-AHW-system-ee}), the lAHW solution becomes self-consistent.
Indeed, the oscillations of $\delta E_{\parallel}$
drive the topological current $\mathbf{J}_{\text{{\tiny AHE}}} \propto[\mathbf{b}\times\delta \mathbf{E}]$, which then induces an oscillating electric field $\delta \tilde{E}_{\perp}$.
Finally, $\delta \tilde{E}_{\perp}$ allows for the component of $\mathbf{J}_{\text{{\tiny AHE}}}$ parallel to $\mathbf{B}_0$,
which together with $\delta E_{\parallel}$ leads to the oscillating CME current,
$\mathbf{J}_{\text{{\tiny CME}}} \propto \mathbf{B}_0 \delta \mu_{5}$, thus closing the cycle.

Despite some similarities with the conventional CMW, which is also
driven by the coupled oscillations of the electric and chiral chemical potentials, the lAHW is a
profoundly different mode that relies on the dynamical electromagnetism and the Chern--Simons
currents in the chiral electron fluid. In this connection, we may add that the CMW is affected by the
dynamical electromagnetism too. As shown in Refs.~\cite{Gorbar:2016ygi,Gorbar:2016sey}
within the CKT, the CMW turns into a chiral plasmon with a nonzero gap.

It is also interesting to briefly discuss the case of Weyl semimetals with broken PI and TR symmetries.
This is covered by including a nonzero $\mu_{5,0}$. At $\mu_{0}=0$, the frequencies of collective modes
affected by $\mu_{5,0}$ are
\begin{eqnarray}
\label{WMHD-B-kz-bx-mu0-omega-gAHW}
\omega_{\text{{\tiny gAHW}},\pm} &\approx& \pm \frac{e^2\sqrt{4e^2b_{\perp}^2g_0 +\varepsilon_e B_0^2 (5\mu_{5,0}^2+7\pi^2T_0^2)}}{\pi \varepsilon_e c \hbar^2 \sqrt{g_0}} +O(k_{\parallel}),\\
\label{WMHD-B-kz-bx-mu0-omega-AHW}
\omega_{\text{{\tiny lAHW}},\pm} &\approx& \pm \frac{B_0 k_{\parallel} \sqrt{(5\mu_{5,0}^2+7\pi^2T_0^2) \left[\pi c^4 \hbar (3\mu_{5,0}^2+\pi^2T_0^2) +3e^4 \mu_{m} v_F^3 b_{\perp}^2\right]}}{c\sqrt{\pi \mu_m \hbar (3\mu_{5,0}^2+\pi^2T_0^2)\left[\varepsilon_e B_0^2(5\mu_{5,0}^2+7\pi^2T_0^2) +4e^2b_{\perp}^2 g_0\right]}} +O(k_{\parallel}^2).
\end{eqnarray}
At the same time, the frequencies of the sound waves $\omega_{\rm s, +}$ and $\omega_{\rm s, -}$ are given by
Eqs.~(\ref{WMHD-B-kz-b0-mu0-omega-s-p}) and (\ref{WMHD-B-kz-b0-mu0-omega-s-m}), respectively.
Similarly to the case $\mathbf{B}_0=\mathbf{0}$, the effect of the chiral chemical potential is quantitative. In the most general case $\mu_0\neq0$ and $\mu_{5,0}\neq0$,
the analytical expressions for the frequencies are too cumbersome to be presented here. They can be easily analyzed,
however, by using numerical methods.

\subsection{Weyl semimetals with $\mathbf{b}\parallel\mathbf{B}_0$}
\label{sec:WMHD-B-kz-bz}

Further, we turn to the case of Weyl semimetals with the chiral shift parallel to the magnetic field,
i.e. $\mathbf{b}\parallel\mathbf{B}_0$. Because of the topological Chern--Simons contribution
in the electric charge density, $\rho_{\text{{\tiny CS}}} \propto \mathbf{b}\cdot\mathbf{B}_0$, the
global equilibrium electric chemical potential $\mu_{0,B}$ is no longer equal to its reference value
at $B_0=0$, i.e. $\mu_0$, see Eq.~(\ref{Minimal-WH-2-compensation-B}).

To start with, let us consider the case where $\mu_0=0$ but a nonzero electric charge density
is topologically induced by the external magnetic field, i.e. $\rho_0=e^3 b_{\parallel}B_0/(2\pi^2 c^2\hbar^2)$.
As in the case $\mathbf{b}\perp\mathbf{B}_0$ and $\mu_0\neq0$, there are eight nontrivial solutions
with rather cumbersome expressions for the dispersion relations.
It is possible to identify among them
the diffusive waves, the gAHW, and the AH mode. The corresponding frequencies read
\begin{eqnarray}
\label{WMHD-B-kz-bz-mu50-neutral-omega-d}
\omega_{\rm d, \pm} &\approx& -\frac{i}{2\tau} \pm \frac{e^3b_{\parallel} \left(4\pi c^2  w_0  +\varepsilon_e v_F^2 B_0^2\right)}{4\pi^2 c^3 \varepsilon_e \hbar^2  w_0 } \nonumber\\
&-& \frac{i \sqrt{\left(2\pi^2 c^3 \varepsilon_e \hbar^2  w_0  \pm4i\pi c^2 e^3  w_0 \tau b_{\parallel} \pm ie^3\varepsilon_e v_F^2 \tau b_{\parallel} B_0^2 \right)^2 \mp 32 i \pi^3 \tau c^5 e^3 \varepsilon_e \hbar^2 b_{\parallel}  w_0 ^2}}{4\pi^2 c^3\varepsilon_e \hbar^2 w_0  \tau} +O(k_{\parallel}),\\
\label{WMHD-B-kz-bz-mu50-neutral-omega-gAHW}
\omega_{\text{{\tiny gAHW}}, \pm} &\approx& -\frac{i}{2\tau} \pm \frac{e^3b_{\parallel} \left(4\pi c^2  w_0  +\varepsilon_e v_F^2 B_0^2\right)}{4\pi^2 c^3 \varepsilon_e \hbar^2  w_0 } \nonumber\\
&+& \frac{i \sqrt{\left(2\pi^2 c^3 \varepsilon_e \hbar^2  w_0  \pm4i\pi c^2 e^3  w_0 \tau b_{\parallel} \pm ie^3\varepsilon_e v_F^2 \tau b_{\parallel} B_0^2 \right)^2 \mp 32 i \pi^3 \tau c^5 e^3 \varepsilon_e \hbar^2 b_{\parallel}  w_0 ^2}}{4\pi^2 c^3\varepsilon_e \hbar^2 w_0  \tau} +O(k_{\parallel}),
\end{eqnarray}
and
\begin{equation}
\label{WMHD-B-kz-bz-mu50-neutrality-omega-h}
\omega_{\text{{\tiny AH}}, \pm} \approx \pm \frac{\pi c^3 \hbar^2k_{\parallel}^2 }{2\mu_m e^3b_{\parallel}} - i\tau\frac{v_F^2B_0^2k_{\parallel}^2 }{4\pi \mu_m  w_0 } +O(k_{\parallel}^3),
\end{equation}
respectively.
Note that the chiral shift has a profound effect on the value of the gap in the gAHW.
In fact, the background magnetic field does not produce any gap by itself when $|\mathbf{b}|=0$. Note also that $\lim_{b_{\parallel}\to0} \omega_{\rm d, \pm}=-i/\tau$, which coincides with that in Eq.~(\ref{WMHD-B0-b0-mu0mu50-omega-d}) obtained for $B_0=0$.
As for the AH modes, their frequencies in Eq.~(\ref{WMHD-B-kz-bz-mu50-neutrality-omega-h}) are somewhat
similar to those in Eqs.~(\ref{WMHD-B0-all-b-mu0mu50-omega-AHW}) and (\ref{WMHD-B0-bz-mu50-omega-h}).

In the case of $\mu_0\neq0$, for the sake of simplicity, we present only the frequencies of the
AH excitations, i.e.
\begin{equation}
\label{WMHD-B-kz-bz-mu50-omega-h}
\omega_{\text{{\tiny AH}}, \pm} \approx - \frac{\pi c^3\hbar^2 \left[ 2i\pi^2\tau c^3 \hbar^2 v_F^2 \rho_0^2 w_0
\mp \left| e^3c^2 b_{\parallel} w_0 ^2 +\tau^2 B_0 v_F^4 \rho_0^2\left(2\pi^2 c^2 \hbar^2 \rho_0 + e^3 b_{\parallel}B_0\right)\right| \right] k_{\parallel}^2}{2\mu_m \left[e^6c^2 b_{\parallel}^2  w_0 ^2+\tau^2v_F^4\rho_0^2\left(2\pi^2c^2\hbar^2 \rho_0 + e^3b_{\parallel} B_0\right)^2\right]} +O(k_{\parallel}^3).
\end{equation}
Note that at $B_0\neq0$ the AH modes in Weyl semimetals contain a nonzero imaginary part, which, in principle, could make them strongly dissipative. For example, by using Eq.~(\ref{WMHD-B-kz-bz-mu50-neutrality-omega-h}), we can estimate the characteristic value of $\tau$, at which the real and imaginary parts are equal, as
\begin{equation}
\label{WMHD-B-kz-bz-mu50-neutrality-omega-h-tau}
\tau^{*} = \frac{2\pi^2c^3\hbar^2  w_0 }{e^3v_F^2 b_{\parallel}B_0^2}.
\end{equation}
For the numerical parameters used previously, see Figs.~\ref{fig:WMHD-B0-AHW}(a) and \ref{fig:WMHD-B0-AHW}(b), as well as Eq.~(\ref{App-realistic-parameters}),
we find that
$\tau^{*} \approx 10^{-12}~\mbox{s}$ and is comparable to typical values of $\tau$ in Weyl materials.

If, in addition to the TR, the PI symmetry is also broken, we find that the frequencies $\omega_{\rm d, \pm}$ and
$\omega_{\text{{\tiny gAHW}}, \pm}$ [see Eqs.~(\ref{WMHD-B-kz-bz-mu50-neutral-omega-d}) and (\ref{WMHD-B-kz-bz-mu50-neutral-omega-gAHW}), respectively] remain unchanged at a nonzero
$\mu_{5,0}$. There are also three gapped solutions and three other solutions with the frequencies quadratic in $k_{\parallel}$.
Their expressions are rather cumbersome and will not be presented here.

\begin{figure}[t]
\begin{center}
\hspace{-0.45\textwidth}(a)\hspace{0.45\textwidth}(b)\\[0pt]
\includegraphics[width=0.45\textwidth]{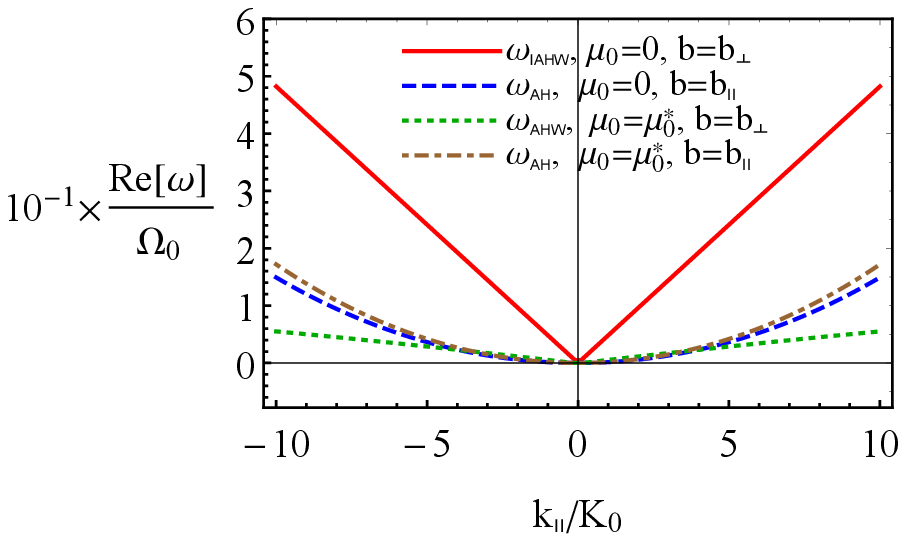} 
\hfill
\includegraphics[width=0.475\textwidth]{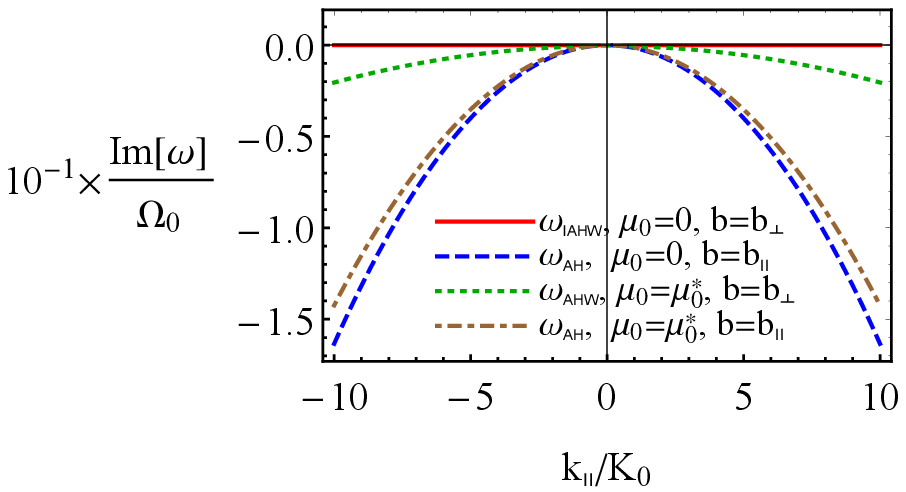}
\caption{The real (panel a) and imaginary (panel b) parts of the AHW and AH frequencies.
Red solid lines correspond to the case $\mathbf{b}\perp\mathbf{k}$ and $\mu_{0}=0$,
blue dashed lines represent $\omega_{\text{{\tiny AH,+}}}$ at $\mathbf{b}\parallel\mathbf{k}$ and $\mu_{0}=0$,
green dotted lines correspond to $\mathbf{b}\perp\mathbf{k}$ and $\mu_{0}=\mu_{0}^{*}$,
brown dot-dashed lines show the results at $\mathbf{b}\parallel\mathbf{k}$ and $\mu_{0}=\mu^{*}_{0}$.
We used the model parameters in Eq.~(\ref{App-realistic-parameters}),
as well as $\mu_{0}^{*}=10~\mbox{meV}$, $\mu_{5,0}=0~\mbox{meV}$, $T_0=10~\mbox{K}$, and $B_0=0.01~\mbox{T}$.
The values of the wave vectors and frequencies are given in units of $K_{0}=10^{-6}\pi/a \approx12.3~\mbox{cm}^{-1}$
and $\Omega_{0}=v_F K_0\approx1.84~\mbox{GHz}$.}
\label{fig:WMHD-B-kz-Helicons}
\end{center}
\end{figure}

The whole spectrum of collective modes can be straightforwardly analyzed by using numerical methods.
The frequencies of the most interesting AHW and AH modes are presented in Figs.~\ref{fig:WMHD-B-kz-Helicons}(a) and \ref{fig:WMHD-B-kz-Helicons}(b).
As in the $B_0=0$ case, in order to avoid the unnecessary clutter, the results for the gAHW are omitted.
The corresponding frequency of the gAHW is determined almost exclusively by the chiral shift, i.e.
$\omega_{\text{{\tiny gAHW}}}\approx
2e^3|\mathbf{b}|/(\pi c \varepsilon_e \hbar^2)$, and depends very weakly
on the wave vector. A small nonzero imaginary part of $\omega_{\text{{\tiny gAHW}}}$ appears at
$\mathbf{b}\parallel\mathbf{B}_0$ and/or at $\mu_0\neq0$. By noting that realistic values of
$\mu_{5,0}$ produce very small changes in the spectra, we set $\mu_{5,0}=0$ for all modes shown
in Fig.~\ref{fig:WMHD-B-kz-Helicons}.

As we see from Figs.~\ref{fig:WMHD-B-kz-Helicons}(a) and \ref{fig:WMHD-B-kz-Helicons}(b), the lAHW is rather strongly affected by the chemical
potential $\mu_0$ and is transformed into a weakly dissipative AHW with a linear dispersion law.
In the case $\mathbf{b}\parallel\mathbf{B}_0$, there are the AH modes with frequencies $\omega_{\text{{\tiny AH}}}
\equiv\omega_{\text{{\tiny AH}}, +}$, which are quadratic in $k_{\parallel}$. Unlike the AHW, they are weakly affected by a
nonzero electric chemical potential $\mu_0$.
By comparing Figs.~\ref{fig:WMHD-B-kz-Helicons}(a) and \ref{fig:WMHD-B-kz-Helicons}(b), we find that the imaginary parts of the AH frequencies are almost of the same order as the real ones.
We conclude, therefore, that such waves are
likely to be highly dissipative or even overdamped in the hydrodynamic regime in Weyl semimetals.

\section{Transverse propagation of collective modes}
\label{sec:WMHD-B-kx}

In this section, we study the transverse collective modes (i.e. $\mathbf{k}\perp\mathbf{B}_0$)
by using the CHD.

\subsection{Dirac semimetals and PI symmetry broken Weyl semimetals}
\label{sec:WMHD-B-kx-b0}

Once again, we start from the simplest case of Dirac semimetals, i.e. $|\mathbf{b}|=b_0=0$.
When the chemical potentials vanish, $\mu_0=\mu_{5,0}=0$, we obtain the usual
types of modes: the diffusive and sound waves, as well as the electromagnetic (light) ones. The
diffusive and light waves have the same frequencies as at $B_{0}=0$, see
Eq.~(\ref{WMHD-B0-b0-mu0mu50-omega-d}) and the text after it.
There is also the gCMW with the frequency similar to that in Eq.~(\ref{WMHD-B-kz-b0-mu0mu50-omega-gCMW}),
albeit with $k_{\parallel}$ replaced by $k_{\perp}$.
As one can see from Eq.~(\ref{WMHD-B-kx-b0-mu0mu50-omega-s-general-k}), the only mode that is truly different is the sound wave.
At $\mu_0\neq0$ (but still at $\mu_{5,0}=0$), the gaps of the diffusive waves and the gCMW
are given by Eqs.~(\ref{WMHD-B-kz-b0-mu50-omega-d}) and (\ref{WMHD-B-kz-b0-mu50-omega-gCMW}), respectively.
The helicons similar to those in Eq.~(\ref{WMHD-B-kz-b0-mu50-omega-h}) are absent, however, when
$\mathbf{k}\perp\mathbf{B}_0$.

By including a nonzero chiral chemical potential, it is also straightforward to generalize the above results to the case of Weyl
semimetals with a broken PI symmetry. When
the electric charge density vanishes, we find in the spectrum the diffusive and sound modes,
the gCMW, and the usual in-medium electromagnetic (light) waves. The gaps for most collective
modes, i.e. the diffusive modes, one of the sound waves, and the gCMW, are given by the same
expressions as in the case of the longitudinal propagation, i.e. Eqs.~(\ref{WMHD-B0-b0-mu0mu50-omega-d}), (\ref{WMHD-B-kz-b0-mu0-omega-s-m})
and (\ref{WMHD-B-kz-b0-mu0-omega-gCMW}), respectively.
The other sound wave, i.e.
$\omega_{\rm s, +}$, is different and its frequency reads
\begin{equation}
\label{WMHD-B-kx-b0-mu0-omega-s-p}
\omega_{\rm s, +} \approx -\frac{i\tau  v_F^2 k_{\perp}^2 \left(15\mu_{5,0}^2+7\pi^2T_0^2\right) \left( w_0  -\mu_{5,0}n_{5,0} -2\sigma^{(\epsilon, u)}B_0^2\right)}{3 w_0  \left(5\mu_{5,0}^2+7\pi^2T_0^2\right)} + O(k_{\perp}^3).
\end{equation}
In the case with $\mu_0\neq0$ and $\mu_{5,0}\neq0$, we find that the frequencies of the diffusive
modes and the gCMW are given by the same expressions as at $\mu_{5}=0$
[see Eqs.~(\ref{WMHD-B-kz-b0-mu50-omega-d}) and (\ref{WMHD-B-kz-b0-mu50-omega-gCMW}), respectively].
There are also three gapped modes with rather
cumbersome dispersion relations.

\subsection{Weyl semimetals with $\mathbf{b}\perp\mathbf{B}_0$}
\label{sec:WMHD-B-kx-all-b-mu0mu50}

In this subsection, we study the collective excitations in a Weyl semimetal with $\mathbf{b}\perp\mathbf{B}_0$.
The results at $\mu_0=\mu_{5,0}=0$ read
\begin{eqnarray}
\label{WMHD-B-kx-all-b-mu0mu50-omega-AHW}
\omega_{\text{{\tiny AHW}},\pm} &=&\pm\frac{1}{c\varepsilon_e\hbar^2 T_0\sqrt{2\pi^3 \mu_m}} \Bigg\{2\pi^3\varepsilon_e \hbar^4 T_0^2 c^4 k_{\perp}^2
+e^4\mu_m\left(4\pi e^2 T_0^2 b_{\perp, \rm tot}^2 +3\varepsilon_e v_F^3\hbar^3B_0^2 \right) \nonumber\\
&-&e^3\sqrt{\mu_m}\sqrt{e^2\mu_m\left(4\pi e^2 T_0^2 b_{\perp, \rm tot}^2 +3\varepsilon_e v_F^3\hbar^3B_0^2 \right)^2 +16\pi^4\varepsilon_e c^4 \hbar^4 T_0^4b_{\perp}^2 k_{\perp}^2} \Bigg\}^{1/2},\\
\label{WMHD-B-kx-all-b-mu0mu50-omega-gAHW}
\omega_{\text{{\tiny gAHW}},\pm} &=&\pm\frac{1}{c\varepsilon_e\hbar^2 T_0\sqrt{2\pi^3 \mu_m}} \Bigg\{2\pi^3\varepsilon_e \hbar^4 T_0^2 c^4 k_{\perp}^2 +e^4\mu_m\left(4\pi e^2 T_0^2 b_{\perp, \rm tot}^2 +3\varepsilon_e v_F^3\hbar^3B_0^2 \right) \nonumber\\
&+&e^3\sqrt{\mu_m}\sqrt{e^2\mu_m\left(4\pi e^2 T_0^2 b_{\perp, \rm tot}^2 +3\varepsilon_e v_F^3\hbar^3B_0^2 \right)^2 +16\pi^4\varepsilon_e c^4 \hbar^4 T_0^4b_{\perp}^2 k_{\perp}^2} \Bigg\}^{1/2},
\end{eqnarray}
where $b_{\perp, \rm tot}^2=b_{\perp}^2+\tilde{b}_{\perp}^2$ (recall that $b_{\perp}$ and $\tilde{b}_{\perp}$ correspond to the components of the chiral shift parallel and perpendicular to $\mathbf{k}$, respectively). The spectrum of collective
modes in this regime also contains the diffusive and sound waves with the same frequencies as in
Eqs.~(\ref{WMHD-B0-b0-mu0mu50-omega-d}) and (\ref{WMHD-B-kx-b0-mu0mu50-omega-s-general-k}),
respectively.

Even at zero charge density, the results for $\omega_{\text{{\tiny AHW}},\pm}$ and $\omega_{\text{{\tiny gAHW}},\pm}$
are rather complicated. In order to get a deeper insight
into the properties of these collective modes, it is useful to consider the two limiting cases:
$\tilde{b}_{\perp}\to0$ and $b_{\perp}\to0$. In the first case, to the leading order in the wave
vector, we obtain
\begin{equation}
\label{WMHD-B-kx-all-bx-mu0mu50-omega-AHMHDW}
\lim_{\tilde{b}_{\perp}\to0}\omega_{\text{{\tiny AHW}},\pm} \approx
\pm \frac{ck_{\perp}\sqrt{3v_F^3\hbar^3}B_0}{\sqrt{\mu_m\left(4\pi e^2T_{0}^2 b_{\perp}^2 +3\varepsilon_e v_F^3\hbar^3 B_0^2\right)}} +O(k_{\perp}^2)
\end{equation}
and the gap $\lim_{\tilde{b}_{\perp}\to0}\omega_{\text{{\tiny gAHW}},\pm}$ is given by Eq.~(\ref{WMHD-B-kz-bx-mu0mu50-omega-gCMW}).
Therefore, when $\mathbf{b}\parallel\mathbf{k}$, the AHW with $\mathbf{k}\perp\mathbf{B}_0$ is transformed
into the \emph{transverse anomalous Hall wave} (tAHW), which was first obtained in Ref.~\cite{Gorbar:2017vph}.
By following the discussion in Ref.~\cite{Gorbar:2017vph}, it is instructive to illustrate the origin of the
tAHW. The corresponding set of nontrivial equations reads
\begin{eqnarray}
\label{WMHD-B-kx-all-bx-mu0mu50-AHMHDW-system-be}
&&\frac{4\epsilon_0 \omega}{T_0} \delta T - k_{\perp}\left( w_0 -2B_0^2\sigma^{(\epsilon, u)}\right) \delta u_{\perp} =0,\\
\label{WMHD-B-kx-all-bx-mu0mu50-AHMHDW-system-2}
&&\frac{k_{\perp}}{T_0} \delta T - \frac{i+\omega\tau}{v_F^2\tau} \delta u_{\perp}
+\frac{5ck_{\perp}^2B_0 \sigma^{(\epsilon, u)}}{\omega  w_0 } \delta \tilde{E}_{\perp} =0,\\
&&\frac{T^2 \omega}{3v_F^3\hbar}\delta\mu_5 -i \frac{e^2B_0}{2\pi^2c} \delta E_{\parallel} =0,\\
&&\left(\omega^2  -\frac{c^2k^2_{\perp}}{\varepsilon_e\mu_m}\right)\delta \tilde{E}_{\perp}
- \frac{2i e^3\omega b_{\perp}}{\pi c\varepsilon_e \hbar^2}\delta E_{\parallel}=0,\\
&&\left(\omega^2  -\frac{c^2k^2_{\perp}}{\varepsilon_e\mu_m}\right)\delta E_{\parallel}
+\frac{2ie^2\omega }{\pi c\varepsilon_e \hbar^2}\left(B_0\delta \mu_5 +eb_{\perp}\delta \tilde{E}_{\perp}\right) =0.
\label{WMHD-B-kx-all-bx-mu0mu50-AHMHDW-system-ee}
\end{eqnarray}
They are the energy conservation, Euler, chiral charge continuity, and two Maxwell's equations.
In the derivation, we also used $\delta \mathbf{B}=(c/\omega) [\mathbf{k}\times \delta \mathbf{E}]$.

Just like in the case of the lAHW, the propagation of the tAHW is sustained by a dynamical version
of the AHE. However, there is an important difference between these two forms of the collective excitations. First, we note that there are oscillations of the electric chemical potential in the former, which are absent in the latter.
Most importantly, while the lAHW is a purely electromagnetic wave, in which the hydrodynamic sector is decoupled, the
tAHW involves both sectors coupled via the magnetic field $\mathbf{B}_0$. Indeed, in the
electromagnetic sector, owing to the chiral anomaly, the oscillating chemical potential $\delta \mu_5$
leads to the electric field $\delta E_{\parallel}$. Then, the AHE allows for the oscillations of $\delta \tilde{E}_{\perp}$
that, in turn, change the local momentum density of the electron fluid via the last term on the left-hand side of
the Euler equation (\ref{WMHD-B-kx-all-bx-mu0mu50-AHMHDW-system-2}). In the end, therefore, the tAHW
is a hybrid collective excitation of the electromagnetic field and the chiral electron fluid that is made possible by the
topological Chern--Simons current.
It is interesting to note that while the tAHW drives oscillations of the fluid velocity, surprisingly, it
remains nondissipative. We can speculate that the presence of an intrinsic viscosity of the electron fluid
might change this nondissipative character.

In the case of $\mathbf{b}\perp\mathbf{k}$, the AHW at $\mathbf{k}\perp\mathbf{B}_0$ is transformed
into the electromagnetic (light) wave $\lim_{b_{\perp}\to0}\omega_{\text{{\tiny AHW}}, \pm}=\omega_{\rm light,\pm}$.
As for the gapped mode, i.e. the gAHW, its frequency becomes
\begin{equation}
\label{WMHD-B-kx-all-by-mu0mu50-omega-gAHW}
\lim_{b_{\perp}\to0}\omega_{\text{{\tiny gAHW}}, \pm} = \pm \frac{\sqrt{\pi^3 c^4\varepsilon_e \hbar^4 T_{0}^2k_{\perp}^2 + 3v_F^3 \varepsilon_e e^4 \hbar^3 B_0^2 +4\pi e^6 \mu_m T_{0}^2\tilde{b}_{\perp}^2}}{c\varepsilon_e \hbar^2 T_{0} \sqrt{\pi^3\mu_m}}.
\end{equation}

Before concluding this subsection, let us also discuss Weyl semimetals in which the PI symmetry
is also broken. In this case $b_0\neq 0$ and the chiral chemical potential is nonzero. There are two
qualitatively different cases with $\mathbf{b}$ parallel and perpendicular to $\mathbf{k}\perp\mathbf{B}_0$,
respectively.

When $\mathbf{b}$ is parallel to $\mathbf{k}$ (and $\rho_0=0$), we find that the frequencies of the diffusive modes, one of the sound modes, and the gAHW
are the same as for the case of the longitudinal propagation of the collective
modes. They are given by Eqs.~(\ref{WMHD-B0-b0-mu0mu50-omega-d}), (\ref{WMHD-B-kz-b0-mu0-omega-s-m}), and (\ref{WMHD-B-kz-bx-mu0-omega-gAHW}). respectively. The frequency of the other sound node, $\omega_{\rm s, +}$, is given by
Eq.~(\ref{WMHD-B-kx-b0-mu0-omega-s-p}). The dispersion relation of the tAHW is
\begin{equation}
\label{WMHD-B-kx-bx-mu0-omega-tAHW}
\omega_{\text{{\tiny tAHW}},\pm} \approx \pm \frac{ck_{\perp}B_0 \sqrt{5\mu_{5,0}^2 +7\pi^2T_0^2}}{\sqrt{\mu_m\left[\varepsilon_e B_0^2 (5\mu_{5,0}^2+7\pi^2T_0^2) +4e^2b_{\perp}^2 g_0\right]}} +O(k_{\perp}^2).
\end{equation}
The same results are also valid for the diffusive modes, the gAHW, and one of the sound modes when $\mathbf{b}\parallel\left[\mathbf{B}_0\times\mathbf{k}\right]$ (and $\rho_0=0$), albeit in the gAHW, one should replace $b_{\perp}\to \tilde{b}_{\perp}$ and the tAHW is replaced by the usual
electromagnetic (light) wave.
The second completely diffusive sound wave has the following frequency:
\begin{equation}
\label{WMHD-B-kx-by-mu0-omega-s-p}
\omega_{\rm s, +} \approx -i\tau \frac{v_F^2k_{\perp}^2 \left[\varepsilon_e B_0^2(15\mu_{5,0}^2+7\pi^2T_0^2) \left( w_0  -\mu_{5,0}n_{5,0}-2\sigma^{(\epsilon, u)}B_0^2\right) +4e^2\tilde{b}_{\perp}^2 g_0 \left( w_0 -2\sigma^{(\epsilon, u)}B_0^2\right)\right] }{3 w_0  \left[\varepsilon_e B_0^2 (5\mu_{5,0}^2+7\pi^2T_0^2) +4e^2\tilde{b}_{\perp}^2 g_0\right]}
+O(k_{\perp}^3).
\end{equation}
In both cases (with $\mathbf{b}\perp\mathbf{B}_0$), at nonzero electric and chiral charge densities, we find that there
is the gapless diffusive wave
with the same frequency as in Eq.~(\ref{WMHD-B0-b0-mu50-omega-d12}).
In addition, there also are seven gapped solutions with rather complicated dispersion relations.

\begin{figure}[t]
\begin{center}
\hspace{-0.45\textwidth}(a)\hspace{0.45\textwidth}(b)\\[0pt]
\includegraphics[width=0.45\textwidth]{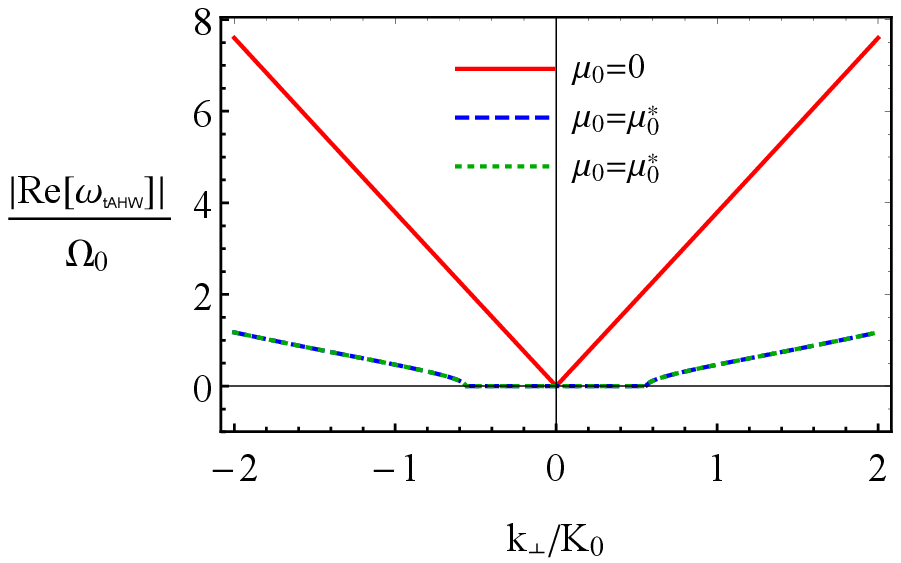}
\hfill
\includegraphics[width=0.475\textwidth]{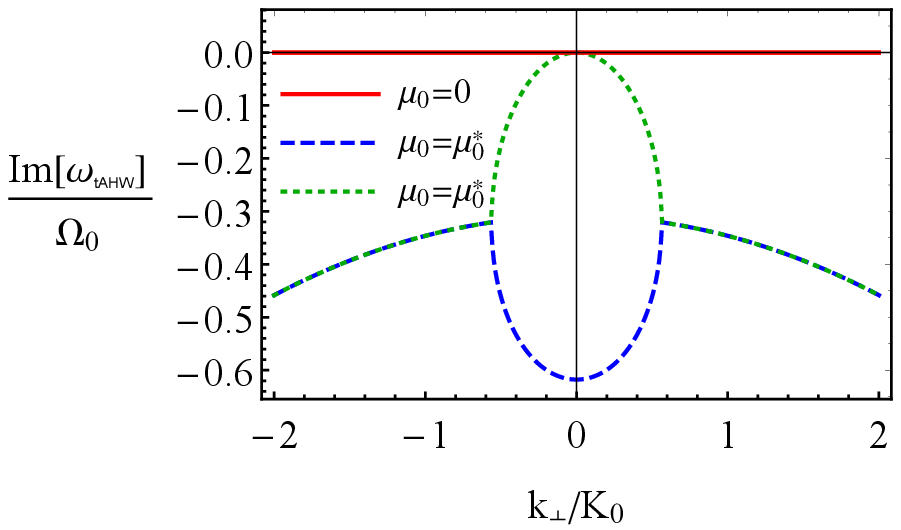}
\caption{The real (panel a) and imaginary (panel b) parts of the tAHW frequency. While red solid lines correspond to
the case $\mu_{0}=0$, blue dashed and green dotted lines represent positive and negative frequencies at
$\mu_{0}=\mu_{0}^{*}=10~\mbox{meV}$. We used the model parameters in Eq.~(\ref{App-realistic-parameters}), as well as $\mu_{5,0}=0$, $T_0=10~\mbox{K}$, and $B_0=0.01~\mbox{T}$.
The values of the wave vectors and frequencies are given in units of $K_{0}=10^{-6}\pi/a \approx12.3~\mbox{cm}^{-1}$
and $\Omega_{0}=v_F K_0\approx1.84~\mbox{GHz}$.}
\label{fig:WMHD-B-kx-tAHW}
\end{center}
\end{figure}

The numerical results for the tAHW frequency, as well as its counterparts at $\mu_0=10~\mbox{meV}$
are presented in Figs.~\ref{fig:WMHD-B-kx-tAHW}(a) and \ref{fig:WMHD-B-kx-tAHW}(b). As in the previous cases, we
do not show the frequency of the gAHW, whose real part is approximately
given by $\omega_{\text{{\tiny gAHW}}}\approx2e^3|\mathbf{b}|/(\pi c \varepsilon_e \hbar^2)$ and is weakly
dispersive.
It also has a small imaginary part that is present at nonzero $\rho_0$ (induced by either $\mu_0$
or $b_{\parallel}$).

As we can see from Figs.~\ref{fig:WMHD-B-kx-tAHW}(a) and \ref{fig:WMHD-B-kx-tAHW}(b), while the frequency of the tAHW is linear in
$k_{\perp}$ and does not contain any imaginary part at $\mu_0=0$, the situation changes
qualitatively at $\mu_0\neq0$.
Indeed, the tAHWs transform into completely diffusive waves at $\mu_0\neq0$ and $|k_{\perp}|\lesssim K_0/2$, see Fig.~\ref{fig:WMHD-B-kx-tAHW}(b).

\subsection{Weyl semimetals with $\mathbf{b}\parallel\mathbf{B}_0$}
\label{sec:WMHD-B-kx-bz}

Finally, we consider the special case of the transverse collective modes (with $\mathbf{k}\perp\mathbf{B}_0$)
when $\mathbf{b}\parallel\mathbf{B}_0$. In the absence of the reference chemical potential, i.e. $\mu_0=0$, the
electric charge density is still induced via the topological contribution, which according to Eq.~(\ref{Minimal-WH-2-compensation-B}),  reads $\rho_0 = e^3\left(\mathbf{b}\cdot\mathbf{B}_0\right)/(2\pi^2 c^2\hbar^2)$.
By solving the characteristic
equation, we find that, to the leading order in $|\mathbf{k}|$, the frequencies of the diffusive waves and the gAHW are given by Eqs.~(\ref{WMHD-B-kz-bz-mu50-neutral-omega-d}) and (\ref{WMHD-B-kz-bz-mu50-neutral-omega-gAHW}), respectively.
The analytical expressions for the frequencies of other collective modes are more complicated.

If both TR and PI symmetries are broken in a Weyl semimetal, we find seven gapped modes at $\mu_0=0$, as well as the following gapless
diffusive one:
\begin{eqnarray}
\label{WMHD-B-kx-bz-mu0-omega-s-p}
\omega_{\rm s, +} &=& -i\tau\frac{v_F^2k_{\perp}^2}{12\pi \mu_m  w_0  (5\mu_{5,0}^2+7\pi^2T_0^2)}
\Bigg\{4\pi \mu_m (15\mu_{5,0}^2+7\pi^2T_0^2) \left( w_0  -\mu_{5,0}n_{5,0}\right) \nonumber\\
&+&B_0^2 \left[15\mu_{5,0}^2 +21\pi^2T_0^2 +4\pi\mu_{m} \left(45\mu_{5,0}^2 +91\pi^2T_0^2\right)\sigma^{(\epsilon, u)} \right]\Bigg\}.
\end{eqnarray}
The dispersion relations of all modes become rather complicated at $\mu_0\neq0$ and $\mu_{5,0}\neq0$
and, consequently, we will not present the corresponding analytical solutions.

\section{Discussions}
\label{sec:Discussions}

In this section, we discuss the range of validity of the present study, as well as its possible experimental
implications. One of the simplifications that we relied on was the use of a Weyl semimetal model with only two Weyl
nodes. The majority of the experimentally discovered (see, e.g. Ref.~\cite{Armitage-Vishwanath:2017-Rev}) Weyl semimetals, however, have multiple pairs of Weyl nodes separated by
$2\mathbf{b}^{(n)}$, where index $n$ labels different pairs of the nodes.
By taking into account that the chiral
shift enters the CHD equations only linearly via the Chern--Simons terms (i.e. the topological charge
and current densities) in Maxwell's equations, the effects of multiple pairs simply add up and can be rendered via $\mathbf{b}\to \mathbf{b}_{\rm eff} \equiv \sum_{n} \mathbf{b}^{(n)}$.
In other words, when the TR symmetry is broken, the spectrum of collective modes is affected  in the same way as predicted
by the model with a single pair of Weyl nodes, albeit with $\mathbf{b}$ replaced by $\mathbf{b}_{\rm eff}\neq 0$.

The situation is different in the special case of Weyl semimetals with an intact TR symmetry, in which
the total number of Weyl nodes is a multiple of four. This is due to the fact that the TR symmetry
maps each pair of the opposite-chirality Weyl nodes separated by $2\mathbf{b}^{(n)}$ to the other pair
separated by $-2\mathbf{b}^{(n)}$. Consequently, the effective chiral shift $\mathbf{b}_{\rm eff}$
should vanish and play no role in the hydrodynamic regime of the TR symmetric Weyl semimetals.
In addition, some observational signatures due to a nonzero chiral chemical potential
$\mu_{5,0}=eb_0\neq 0$ should be still expected if the PI symmetry is broken.

Another simplification of the present study was the treatment of the dissipation effects.
In particular, we used the relaxation-time approximation in order to estimate the chirality-preserving (intravalley)
scattering processes (with the relaxation time $\tau$), but completely ignored the chirality-flipping
(intervalley) processes (with the relaxation time $\tau_5$). Such an approximation can be
justified by noting that usually $\tau_5\gg \tau$ (see, e.g. Ref.~\cite{Zhang-Xiu:2015}).
In addition, the dependence of the relaxation time on the chemical potentials and temperature was also ignored. While
such a dependence leads only to quantitative corrections to the quasiparticles dispersion relations,
it could provide qualitative effects in the dependence of $\omega$ on $\mu_0$ and/or $T_0$.

We would like to argue that many predicted distinctive features of the collective modes spectrum
that stem from the chiral shift
can be observed in realistic materials. Among the most promising candidates,
we could suggest the antiferromagnetic half Heusler compounds, such as GdPtBi and NdPtBi. The latter
were predicted to be TR symmetry breaking Weyl semimetals \cite{Hirschberger-Ong:2016,Shekhar-Felser:2016,Suzuki:2016}
in an applied magnetic field. Another type of suitable Weyl semimetals are the theoretically proposed magnetic
Heusler compounds such as XCo$_2$Z (where X=V,Zr,Ti,Nb,Hf and Z=Si,Ge,Sn), VCo$_2$Al, and VCo$_2$Ga
\cite{Hasan-magnetic:2016,Cava-Bernevig:2016}. As an added benefit, the latter have only two Weyl nodes near
the Fermi level and the separation between the nodes is comparable to the size of the Brillouin zone.

In order to experimentally test the unusual features of the collective excitations in Weyl semimetals,
one could use a similar experimental setup as in usual metals (see, e.g. Ref.~\cite{Maxfield}). This
requires, in particular, measuring the transmission amplitude of electromagnetic waves through a Weyl
crystal as a function of the frequency at a fixed external magnetic field or as a function of the field at
a fixed frequency. The response will be a series of resonances originating from the interference of
standing waves, from which the parameters of the dispersion laws could be extracted. Depending
on an experimental technique, the effects of various directions of the chiral shift can be studied
by changing the orientation of the magnetic field and/or crystal.

\section{Summary}
\label{sec:Summary}

In this paper, by making use of the consistent hydrodynamic framework proposed in Ref.~\cite{Gorbar:2017vph},
we analyzed the spectrum of collective excitations in Weyl and Dirac materials
with and without a background magnetic field. The underlying framework includes the effects of the chiral
anomaly and the vorticity, as well as the topological Chern--Simons (or Bardeen--Zumino) contributions to the
electric current and charge densities.
While the charge and current densities in Maxwell's equations include the topological Chern--Simons terms,
no such terms appear in the Euler equation and the energy conservation
relation. This can be explained by the fact that the hydrodynamic
sector of the theory is not
sensitive to the filled electron states deep below the Fermi surface, which are primarily responsible for the
Chern--Simons terms. At the same time, Maxwell's equations are affected by the total charge
and current densities, including those from the filled electron states.
As a result, the properties of collective modes are profoundly modified by $b_0$ and $\mathbf{b}$,
which determine the separations between the Weyl nodes in energy and momentum, respectively.

By considering the limit of vanishing magnetic field and electric charge density,
we found that the spectrum of collective excitations in Dirac semimetals
includes a diffusive wave, hydrodynamic sound modes, and in-medium electromagnetic
(light) waves. Generically, the presence of an electric charge density $\rho_0$ modifies these waves
by generating nonzero gaps. When the external magnetic field $\mathbf{B}_0$ is present, the gaps are
induced for the electromagnetic waves even in the absence of an electric charge density. In addition,
we also found that a nonzero $\mathbf{B}_0$ changes the dispersion relation of the sound wave.
Further, in agreement with the findings in Ref.~\cite{Pellegrino}, the external magnetic field gives rise to helicons
propagating along the direction of $\mathbf{B}_0$.
In the presence of the electron-phonon interaction and/or the electron scattering on impurities, most of the collective modes,
including the helicons, become dissipative.
This is not true, however, for the modes that are primarily electromagnetic in nature.

In Weyl semimetals with a broken TR symmetry, we found that the topological contributions associated with
the chiral shift parameter qualitatively change the properties of some collective modes and, in some cases,
give rise to novel types of excitations. For example, in the case of vanishing background magnetic field
and zero electric charge density, the presence of the chiral shift $\mathbf{b}$ produces a gap proportional to
$|\mathbf{b}|$ for one of the anomalous Hall waves. As for the other (gapless) anomalous Hall wave, depending on the
orientation of the wave vector $\mathbf{k}$ with respect to $\mathbf{b}$, it interpolates between the usual
in-medium electromagnetic (light) wave (at $\mathbf{k} \perp\mathbf{b}$) and the anomalous helicon wave
(at $\mathbf{k} \parallel\mathbf{b}$). The latter is a unique helicon with a quadratic dispersion relation that
appears in Weyl semimetals in the absence of background magnetic fields and at vanishing electric charge density.

By studying the spectra of longitudinal and transverse (with respect to the direction of $\mathbf{B}_0$) collective waves,
it is found that $\mathbf{B}_0$ strongly affects the corresponding dispersion relations.
At vanishing electric charge density and $\mathbf{b}\perp\mathbf{B}_0$, the longitudinal modes include diffusive waves, sound modes, and the anomalous Hall waves.
One of these Hall waves is gapped and the value of its gap is determined by the
magnetic field and the chiral shift. Another closely related excitation is a special type of the anomalous Hall waves called the
\emph{longitudinal anomalous Hall wave}. This dissipationless wave
involves the oscillations of the electric and chiral charge densities as well as the electric field perpendicular
to the chiral shift. A dynamical version of the anomalous Hall effect plays an essential role in sustaining
the propagation of such a wave. In the case of $\mathbf{b}\parallel\mathbf{B}_0$, we found modified
gapped anomalous Hall and diffusive waves, as well as the anomalous helicons. While the latter are somewhat
similar to those at $\mathbf{B}_0=\mathbf{0}$, they are dissipative. Our numerical estimates suggest that,
for realistic values of model parameters, the dissipation may be sufficiently strong to completely damp the
helicon waves.

Generically, the transverse collective modes at vanishing electric charge density are of the
same type: diffusive waves, sound modes, and the anomalous Hall waves. When the wave vector is parallel
to the chiral shift, one of anomalous Hall waves turns into the \emph{transverse anomalous Hall wave} that was previously
discussed in Ref.~\cite{Gorbar:2017vph}. When $\rho_0=0$, its frequency is real, linear in the wave vector, as well as
proportional to the magnetic field and inversely proportional to the chiral shift. While this mode looks similar to
the longitudinal anomalous Hall wave, it is quite different in nature. Indeed, this is a hybridized
version of electromagnetic and hydrodynamic waves. At $\rho_0\neq 0$, we found that the transverse anomalous
Hall wave is transformed into a dissipative wave with a quadratic dispersion law. At sufficiently small values of
the wave vector, this wave becomes completely diffusive and resembles the behavior of an overdamped
sound mode.

\begin{acknowledgments}
The work of E.V.G. was partially supported by the Program of Fundamental Research of the
Physics and Astronomy Division of the National Academy of Sciences of Ukraine.
The work of V.A.M. and P.O.S. was supported by the Natural Sciences and Engineering Research Council of Canada.
The work of I.A.S. was supported by the U.S. National Science Foundation under Grants PHY-1404232
and PHY-1713950.
\end{acknowledgments}

\appendix

\section{Additional formulas}
\label{sec:app-formulas}

In this section, we present the explicit expressions for the anomalous transport coefficients and some anomalous Hall waves (AHW) frequencies.

\subsection{Anomalous transport coefficients}
\label{sec:app-formulas-coeff}

The anomalous transport coefficients used in the consistent hydrodynamics equations are given by
\begin{eqnarray}
\label{model-sigma-CKT-be}
\sigma^{(B)} &=& \frac{e^2\mu_5}{2\pi^2\hbar^2c}, \qquad \sigma_5^{(B)} = \frac{e^2\mu}{2\pi^2\hbar^2c}, \\
\sigma^{(\epsilon, B)} &=& -\frac{e\mu\mu_5}{2\pi^2\hbar^2v_F c}, \qquad \sigma^{(\epsilon, u)} = \frac{e^2v_F}{120\pi^2\hbar c^2},\\
\sigma^{(V)}&=&-\frac{e\mu\mu_{5}}{\pi^2v_F^2\hbar^2}, \qquad \sigma_5^{(V)} = -\frac{e}{2\pi^2\hbar^2v_F^2}\left(\mu^2+\mu_5^2 + \frac{\pi^2T^2}{3}\right),\\
\sigma^{(\epsilon, V)} &=& -\frac{e\mu}{6\pi^2\hbar v_F}, \qquad \sigma^{(\epsilon, V)}_5 = -\frac{e\mu_5}{6\pi^2\hbar v_F},
\label{model-sigma-CKT-ee}
\end{eqnarray}
where $e$ is the absolute value of the electron charge, $\mu$ denotes the electric chemical potential, $\mu_5$ is the chiral chemical potential, $T$ is temperature, $v_F$ is the Fermi velocity, and $c$ in the speed of light.
Note that the coefficients in Eqs.~(\ref{model-sigma-CKT-be})--(\ref{model-sigma-CKT-ee}) agree with those obtained
in Refs.~\cite{Son:2012wh,Landsteiner:2012kd,Stephanov:2015roa} in the ``no-drag" frame
\cite{Rajagopal:2015roa,Stephanov:2015roa,Sadofyev:2015tmb}.

\subsection{Anomalous Hall waves frequencies}
\label{sec:app-formulas-frequencies}

Next, we present the frequencies of some AHW. Let us start with the case of the Weyl semimetal with a broken time-reversal symmetry in the absence of the external magnetic field $\mathbf{B}_0$. The frequencies of the AHWs in the leading order in the wave vector $\mathbf{k}$ are
\begin{eqnarray}
\label{WMHD-B0-all-b-mu50-omega-AHW}
\omega_{\text{{\tiny AHW}}, \pm} &\approx& -\frac{i w_0 |\mathbf{k}|^2}{24\pi \mu_m v_F^2\rho_0^2\tau \left(4\pi^4 c^2v_F^4 \hbar^4 \rho_0^4 \tau^2 +e^6 w_0^2|\mathbf{b}|^2\right)} \Bigg\{
e^6b_{\perp}^2 w_0\left(3c^2w_0+4\pi \mu_m v_F^4\rho_0^2\tau^2\right) +24 \pi^4 v_F^4c^4 \hbar^4 \rho_0^4\tau^2\nonumber\\
&\pm& e^3 w_0 \sqrt{e^6b_{\perp}^4\left(3c^2 w_0  -4\pi \mu_m v_F^4 \rho_0^2 \tau^2\right)^2 -48\pi c^2v_F^4 \rho_0^2 \tau^2 b_{\parallel}^2 \left(3\pi^3c^4\hbar^4 \rho_0^2 +e^6\mu_m b_{\perp}^2 w_0 \right)} \Bigg\} +O(|\mathbf{k}|^3).
\end{eqnarray}
Here $w_0=\epsilon_0+P_0$ is the equilibrium enthalpy density, $\epsilon_0$ is the equilibrium energy density, $P_0$ is the equilibrium pressure, $\rho_0$ is the equilibrium electric charge density, $\mu_m$ is the magnetic permeability, $\varepsilon_e$ is the electric permittivity, and $\tau$ is the relaxation time connected with the intravalley (chirality preserving) scattering. [See also Eqs.~(\ref{Minimal-WH-2-equilibrium-be})--(\ref{Minimal-WH-2-equilibrium-ee}) and the notations after them in the main text of the paper.] In addition, we used $b_{\parallel}$ and $b_{\perp}$ for the parallel and perpendicular components of the chiral shift defined with respect to the direction of the wave vector $\mathbf{k}$.

Further, we consider the case of the longitudinal, with respect to the external magnetic field $\mathbf{B}_0$, propagation of the collective waves.
In addition, we assume that the chiral shift is perpendicular to the magnetic field, i.e. $\mathbf{b}\perp\mathbf{B}_0$ and $|\mathbf{b}|=b_{\perp}$.
The frequencies of the gapped anomalous Hall wave (gAHW) and longitudinal anomalous Hall wave (lAHW), which are considered in Sec.~\ref{sec:WMHD-B-kz-bx}, read
\begin{eqnarray}
\label{WMHD-B-kz-bx-mu0mu50-omega-gCMW-app}
\omega_{\text{{\tiny gAHW}},\pm} &=& \pm\frac{1}{2\pi^2 \varepsilon_e c\hbar^2T_0^2\sqrt{2\mu_m}} \Bigg\{4\pi \mu_m e^4 T_0^2 \left(4\pi e^2 T_0^2 b_{\perp}^2 +3\varepsilon_e v_F^3 \hbar^3 B_0^2\right) +\varepsilon_e \hbar^4 k_{\parallel}^2\left(4\pi^4c^4T_0^4 +9\mu_m \varepsilon_e \hbar^2 v_F^6 e^2B_0^2\right) \nonumber\\
&+&\Big\{\varepsilon_e^2\hbar^6\left[4\pi^4 c^4 \hbar T_0^4 k_{\parallel}^2 -3e^2B_0^2 \mu_m v_F^3 \left(4\pi e^2T_0^2+3\varepsilon_e \hbar^3v_F^3 k_{\parallel}^2\right)\right]^2
+32\pi^2\varepsilon_e \mu_m \hbar^3e^6 T_0^4 b_{\perp}^2 \nonumber\\
&\times&\left[4\pi^4c^4\hbar T_0^4 k_{\parallel}^2 +3\mu_mv_F^3 e^2B_0^2\left(4\pi e^2T_0^2-3\varepsilon_e \hbar^3v_F^3 k_{\parallel}^2\right) \right] +256\pi^4\mu_m^2 e^{12} T_0^8 b_{\perp}^4 \Big\}^{1/2} \Bigg\}^{1/2}
\end{eqnarray}
and
\begin{eqnarray}
\label{WMHD-B-kz-bx-mu0mu50-omega-AHW-app}
\omega_{\text{{\tiny lAHW}},\pm} &=& \pm\frac{1}{2\pi^2 \varepsilon_e c \hbar^2T_0^2\sqrt{2\mu_m}} \Bigg\{4\pi \mu_m e^4 T_0^2 \left(4\pi e^2 T_0^2 b_{\perp}^2 +3\varepsilon_e v_F^3 \hbar^3 B_0^2\right) +\varepsilon_e \hbar^4 k_{\parallel}^2 \left(4\pi^4c^4T_0^4 +9\mu_m \varepsilon_e \hbar^2 v_F^6 e^2B_0^2\right) \nonumber\\
&-&\Big\{\varepsilon_e^2\hbar^6\left[4\pi^4 c^4 \hbar T_0^4 k_{\parallel}^2 -3e^2B_0^2 \mu_m v_F^3 \left(4\pi e^2T_0^2+3\varepsilon_e \hbar^3v_F^3 k_{\parallel}^2\right)\right]^2 +32\pi^2\varepsilon_e \mu_m \hbar^3e^6 T_0^4 b_{\perp}^2 \nonumber\\
&\times&\left[4\pi^4c^4\hbar T_0^4 k_{\parallel}^2 +3\mu_mv_F^3 e^2B_0^2\left(4\pi e^2T_0^2-3\varepsilon_e \hbar^3v_F^3 k_{\parallel}^2\right) \right] +256\pi^4 \mu_m^2 e^{12} T_0^8 b_{\perp}^4 \Big\}^{1/2} \Bigg\}^{1/2},
\end{eqnarray}
respectively. Their counterparts in the long-wavelength limit are given by Eqs.~(\ref{WMHD-B-kz-bx-mu0mu50-omega-gCMW}) and (\ref{WMHD-B-kz-bx-mu0mu50-omega-AHW}), respectively, in the main text.

\end{document}